\pdfoutput=1
\documentclass[12pt]{template}
\usepackage{soul}
\usepackage[utf8]{inputenc}
\usepackage[english]{babel}
\usepackage{graphicx}
\usepackage{subcaption}
\usepackage{url}
\usepackage[colorlinks=true, allcolors=blue]{hyperref}

\usepackage{enumitem}
\usepackage{multirow}
\usepackage{amsmath}
\usepackage{comment}
\usepackage{float}
\usepackage{svg}
\usepackage{ulem}
\usepackage{tikz}
\usepackage{longtable}
\usepackage{colortbl} 

\providecommand{\keywords}[1]
{
  \small	
  \textbf{\textit{Keywords---}} #1
}

\usepackage{MnSymbol}

\usepackage{verbatim}
\usepackage{enumitem}
\usepackage{subcaption}

\usepackage{listings}
\usepackage{xcolor}
\usepackage{algorithm}

\usepackage{tcolorbox}
\lstdefinestyle{pseudoStyle}{
    basicstyle=\ttfamily\small,
    numbers=left,
    numberstyle=\tiny\color{gray},
    numbersep=10pt,
    backgroundcolor=\color{white},
    showspaces=false,
    showstringspaces=false,
    showtabs=false,
    frame=single,
    tabsize=2,
    captionpos=b,
    breaklines=true,
    breakatwhitespace=false,
    xleftmargin=0pt,
    framexleftmargin=1.5em,  
    framexrightmargin=0pt,
    framexbottommargin=0.3em
}

\title{Enhancing NeuroEvolution-Based Game Testing: A Branch Coverage Approach for Scratch Programs
}
\author[1,2]{Khizra Sohail}

\author[1]{Atif Aftab Ahmed Jilani }
\author[1]{Nigar Azhar Butt}

\affil[1]{Department of Software Engineering, National University of Computer and Emerging Sciences, Islamabad, Pakistan}

\corrauthor[2]{Khizra Sohail}{khizrasohail739@gmail.com}

\begin{abstract}
Automated test generation for game-like programs presents unique challenges due to their non-deterministic behavior and complex control structures. The NEATEST framework has been used for automated testing in Scratch games, employing neuroevolution-based test generation optimized for statement coverage. However, statement coverage alone is often insufficient for fault detection, as it does not guarantee execution of all logical branches. This paper introduces a branch coverage-based fitness function to enhance test effectiveness in automated game testing.
We extend NEATEST by integrating a branch fitness function that prioritizes control-dependent branches, guiding the neuroevolution process to maximize branch exploration. To evaluate the effectiveness of this approach, empirical experiments were conducted on 25 Scratch games, comparing Neatest with Statement Coverage ($N_{SC}$) against Neatest with Branch Coverage ($N_{BC}$). A mutation analysis was also performed to assess the fault detection capabilities of both techniques. The results demonstrate that $N_{BC}$ achieves higher branch coverage than $N_{SC}$ in 13 out of 25 games, particularly in programs with complex conditional structures. Moreover, $N_{BC}$ achieves a lower false positive rate in mutation testing, making it a more reliable approach for identifying faulty behavior in game programs.
These findings confirm that branch coverage-based test generation improves test coverage and fault detection in Scratch programs. 

\end{abstract}

\begin{document}







\flushbottom
\maketitle
\thispagestyle{empty}

\section{Introduction}
\label{intro}
The gaming industry is a rapidly growing sector, with an estimated net worth of US\$ 522.46 billion worldwide in 2025, experiencing an annual growth rate of 7.25\% \cite{statista2025games}. Games are intentionally designed to be interactive, unpredictable, and engaging, which makes automated testing a challenging task \cite{deiner2023automated}. Unlike traditional software applications, games rely on complex user interactions and randomized events, making it difficult to apply conventional test generation strategies \cite{politowski2021survey}. Even when a test generator successfully identifies a relevant sequence of actions, the randomized nature of game execution makes it nearly impossible to reproduce the same test case in subsequent runs. This non-deterministic behavior complicates test generation, as traditional static test suites cannot adapt to dynamically evolving game states.

Currently, games can be tested using manual testing, semi-automated testing, or fully automated testing. Manual Testing  is widely used in the gaming industry but is labor-intensive and relies on human expertise \cite{diah2010usability}, \cite{ferre2009playability}, \cite{politowski2021survey}. Semi-Automated Testing combines manual and automated approaches, reducing workload but still requiring human intervention \cite{cho2010online}, \cite{schaefer2013crushinator}, \cite{smith2009computational}. Automated Testing, in contrast, aims to generate and execute test cases dynamically, offering a scalable and efficient alternative \cite{ariyurek2019automated}, \cite{pfau2017automated}, \cite{politowski2022towards}.
Traditional automated techniques often either record and replay test scenarios \cite{xue2022learning} or use recorded test cases as base for evolutionary search \cite{feldmeier2023learning}, but due to the inherent randomness in games, these static test suites fail to provide adequate coverage. Dynamic test suites, which react to unseen game states and adjust test execution in real-time, are necessary to improve the effectiveness of automated testing.

Scratch \cite{maloney2010scratch} is a widely used block-based programming language with over 90 million registered users, primarily used for introducing programming concepts to beginners. Many Scratch projects involve game-like elements, providing a learning environment that fosters problem-solving and creativity. Despite its simplicity, Scratch games can contain logical errors that require systematic testing. The Whisker framework \cite{stahlbauer2019testing} offers an execution environment for automated test execution of Scratch programs, supporting both manually written and automatically generated test cases. However, while an execution framework exists, a critical challenge remains: how to generate test suites automatically in a way that ensures adequate test coverage and fault detection.

NEATEST \cite{feldmeier2022neuroevolution} is an automated test suite generator for Scratch games, leveraging NeuroEvolution of Augmenting Topologies (NEAT) \cite{stanley2002evolving} to evolve dynamic test agents in the form of neural networks. These neural networks act as test input generators, dynamically interacting with the game environment to reach specific program states. Test generation in NEATEST is by default guided by statement coverage criteria. However, statement coverage alone is one of the weakest coverage criteria, as it does not ensure that all logical branches of a program are tested.

\begin{figure}
  \centering
  \includegraphics{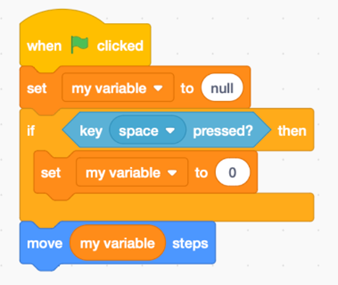}  
 \center \caption{Code Sample from Scratch}
  \label{fig:scratch-code}
\end{figure}

In programs with control structures, execution follows different branches based on decision-making conditions. A fault located in a specific branch may remain undetected if that branch is never executed. NEATEST’s statement coverage-based test generation does not guarantee the execution of all logical branches, making it insufficient for uncovering certain types of faults. Consider the example in Figure \ref{fig:scratch-code}, where a test suite achieves 100\% statement coverage but fails to detect a fault because the alternate branch of an if-condition is never triggered. This limitation underscores the need for a more robust coverage strategy that prioritizes branch execution.

Several coverage criteria exist in structural testing, but selecting the appropriate default criterion for an automated test generator is a challenging task. Coverage selection is often based on industry standards or optimized for specific testing goals, such as improving fault detection rates. Branch coverage is widely considered a more effective test coverage metric than statement coverage, as it ensures that both true and false branches of conditional statements are exercised.

For a general-purpose test generator, branch coverage is often preferred because it balances thoroughness and efficiency \cite{fraser2012whole}. It captures logical execution paths, increasing the likelihood of detecting faults in branching logic, without being as computationally expensive as full path coverage\cite{ntafos1979path}, \cite{yang1998all}.

To address the limitations of statement coverage-based testing, this paper introduces a branch coverage-based fitness function for NEATEST. We define and evaluate two coverage criteria along with their respective fitness functions for guiding NeuroEvolution-based test generation. Our contributions are summarized as follows:

\begin{enumerate}
\itemsep0em 
    \item We propose a new coverage strategy, ($N_{BC}$), by developing a branch coverage-based fitness function integrated with NEATEST.
    \item We perform an empirical comparison of coverage percentages achieved by Neatest with Branch Coverage ($N_{BC}$) and Neatest with Statement Coverage ($N_{SC}$) across 25 Scratch games.
    \item We conduct mutation analysis for $N_{BC}$ and $N_{SC}$, evaluating their fault detection effectiveness using mutation kill score percentages.
\end{enumerate}

This work extends NEATEST by integrating branch coverage-based test generation and demonstrates its effectiveness in improving test coverage and fault detection in Scratch games. Through empirical analysis, we show that branch coverage-based fitness functions enable more efficient and effective test case generation, making them a viable alternative to statement coverage-based approaches.

The remainder of this paper is structured as follows: Section \ref{sec:back} provides the background, detailing key concepts and covers related works. Section \ref{sec:approach} describes the proposed methodology. Section \ref{sec:evaluation} outlines the experimental design, research questions, and evaluation metrics. Section \ref{sec:results} presents the results and analysis, discussing the performance of \(N_{BC}\) against \(N_{SC}\). Section \ref{sec:threats} details threats to validity. Finally, Section \ref{sec:conclusion} concludes the paper by summarizing the contributions and highlighting avenues for future research.

\section{Background and Literature Review}\label{sec:back}
This section provides an overview of test coverage criteria (statement and branch coverage) and their significance in search-based software testing (SBST) and neuroevolution-based test generation. Additionally, it discusses prior research related to automated game testing, test generation strategies, and Scratch testing frameworks to position this study within the broader context of existing work.

\subsection{Software Testing}
Software testing is a critical component in ensuring the quality, reliability, and correctness of software applications. However, it presents several challenges, including ensuring sufficient test coverage, evaluating test effectiveness, detecting critical faults, and maintaining test cases across evolving software versions. One of the most pressing challenges is the evaluation of test quality \cite{graham2008foundations}, which is typically assessed using test adequacy criteria \cite{hemmati2015effective}, metrics that determine whether a test suite sufficiently exercises the software system.

Among these adequacy criteria, statement coverage remains one of the most widely used measures for evaluating test completeness \cite{zhu1997software}. Test coverage, in particular, ensures that test cases execute different elements of a program’s code, helping to uncover defects. Several studies have examined the effectiveness of coverage-based test adequacy criteria in detecting software faults, emphasizing the importance of selecting an appropriate metric for test generation.

This study specifically focuses on two widely used Test coverage criteria: statement coverage and branch coverage.

\subsection{Test Coverage Criteria}
Test coverage measures the extent to which a test suite exercises a program’s control flow. The most commonly used coverage criteria in search-based testing are statement coverage and branch coverage.

\subsubsection{Statement Coverage}
A statement is the smallest executable unit of a program, ranging from single-line instructions to multi-line compound statements. Achieving 100\% statement coverage ($SC$) implies that every executable statement in the program has been executed at least once \cite{graham2008foundations}. $SC$ is a metric used to measure the degree to which the source code of a program is executed when a particular test suite is run in terms of the number of lines of code (LOC) executed versus total LOC as shown in equation \ref{eq:codecoverage}.

\begin{equation}\label{eq:codecoverage}
SC = \frac{\text{LOC executed}}{\text{Total LOC}} \times 100\%
\end{equation}

Statement coverage is one of the simplest and most commonly used test adequacy criteria, particularly in industry applications. However, it does not guarantee comprehensive fault detection since it does not consider whether all control flow paths have been executed. Prior research has identified statement coverage as one of the weakest test coverage metrics, as it may leave critical logic paths untested \cite{zhu1997software}, \cite{hemmati2015effective}. Despite its limitations, it remains widely used due to its ease of implementation and computational efficiency.

\subsubsection{Branch Coverage}
Branch coverage is a stronger structural testing criterion that ensures all control flow branches (i.e., both true and false conditions of decision statements) are executed at least once during testing. It is measured as the ratio of executed branches to total branches in the program (as shown in equation \ref{eq:decisioncoverage}), providing a more thorough assessment of test adequacy compared to statement coverage \cite{graham2008foundations}.

\begin{equation}\label{eq:decisioncoverage}
BC = \frac{\text{Branches executed}}{\text{Total Branches}} \times 100\%
\end{equation}

Since branch coverage inherently subsumes statement coverage, achieving 100\% branch coverage guarantees 100\% statement coverage, but not vice versa . Several studies have demonstrated that branch coverage-based testing improves fault detection rates, as it ensures that decision points within the code are properly evaluated \cite{zhu1997software}, \cite{hemmati2015effective}.

Consider the pseudo-code example (Listing \ref{lst:openacc}): If a test input \texttt{BAL = 6000} is used, 100\% statement coverage is achieved, but only 50\% branch coverage, as the false branch of the if-condition is never executed. A hidden fault (returning a NULL reference) remains undetected unless a second test case, such as \texttt{BAL = 4000}, is executed. This second test case enables full branch coverage, ensuring that both execution paths are tested. This example highlights why branch coverage is a superior metric for structural testing.

\vspace{0.5em}
\begin{lstlisting}[style=pseudoStyle, caption={Function \texttt{OpenAcc}: Creates a new bank account if the initial balance is at least 5000. Otherwise, returns \texttt{NULL}.}, label={lst:openacc}]
Function BankAcc OpenAcc(int BAL)
    If (BAL >= 5000) Then
        Return new BankAcc(BAL)
    EndIf
    Return NULL
EndFunction
\end{lstlisting}

\subsection{Search-Based Software Testing (SBST) for Test Generation}
Search-Based Software Testing (SBST) \cite{mcminn2004search} employs meta-heuristic search algorithms to generate test inputs that optimize test coverage. These algorithms iteratively refine test cases based on fitness functions, which measure how well a test reaches specific program states.

\subsubsection{Fitness Function in SBST}

The effectiveness of an SBST approach depends on the fitness function guiding the search. Two widely used fitness components are:

\begin{enumerate}
\itemsep0em 
    \item \textbf{Approach Level}: Determines the distance between the executed path and the target statement, encouraging the search process to move toward uncovered code \cite{wegener2001evolutionary}.
    \item \textbf{Branch Distance}: Measures how close a test execution is to reaching a target branch, influencing the selection of test inputs \cite{korel1990automated}.
\end{enumerate}
SBST has been successfully applied in test case generation for method call sequences\cite{baresi2010testful},\cite{fraser2011evosuite}, \cite{tonella2004evolutionary}, REST API testing
\cite{arcuri2019restful}, 
and GUI testing \cite{gross2012search},\cite{mao2016sapienz}, \cite{sell2019empirical}
. However, it has limitations when applied to game-like environments, where static test inputs often fail to account for randomized behavior.

\subsection{NeuroEvolution-Based Test Generation}
To overcome the static nature of SBST-generated test inputs, researchers have explored neuroevolution-based test generation, which evolves dynamic test agents in the form of neural networks. 

Neuroevolution has been applied in multiple domains, including game AI testing and software testing automation.

\subsubsection{NEATEST Framework}

NEATEST \cite{feldmeier2022neuroevolution} extends the Whisker framework \cite{stahlbauer2019testing} to support dynamic test generation for Scratch programs. It uses NeuroEvolution of Augmenting Topologies (NEAT) \cite{stanley2002evolving} to evolve neural networks that act as adaptive test input generators, enabling robust testing despite randomized behaviors in games.

Neatest optimizes for statement coverage by default, prioritizing the sequential execution of program statements.
Target statements are selected using a Control Dependence Graph (CDG) \cite{cytron1991efficiently}, allowing the search to progress systematically through the program.
Fitness functions are used to guide the evolution of neural networks, incorporating both branch distance and approach level.

\subsubsection{Neural Networks as Test Oracles}  
In NEATEST \cite{feldmeier2022neuroevolution}, neural networks function as test oracles by evaluating program execution and detecting erroneous behavior. During test execution, the network monitors game states and responses to determine whether observed behavior aligns with expected execution patterns. NEATEST uses node activations and structural changes in the evolved network to assess deviations.  

If a neural network exhibits unexpected activation patterns when interacting with a mutated version of the program, the mutation is flagged as a fault. Additionally, large deviations in network structure or activation values indicate that the tested program behaves differently than expected, helping to distinguish between genuine faults and normal randomized variations in execution.

\subsubsection{Adapting NEATEST for Branch Coverage}

While previous work on NEATEST has focused on statement coverage, this study introduces a branch coverage-based fitness function to improve fault detection capabilities. By modifying the fitness function, we enable NEATEST to prioritize branch execution instead of merely maximizing statement coverage.
Hence, test diversity can be improved by ensuring both true and false branches of decision statements are executed.
This enhances fault detection rates, as uncovered branches may contain hidden faults.

\subsection{Testing Scratch Programs}
Scratch is a beginner-friendly, block-based programming language that allows users to create interactive games, animations, and stories by assembling visual code blocks. Each project features a stage (background) and sprites (characters or objects), whose behaviors are controlled through block-based scripts. While the visual blocks enforce syntactic correctness, functional errors can still arise, making testing essential. To address this, Whisker automates testing by providing inputs to Scratch programs and checking their behavior against predefined specifications.
Whisker offers two types of test harnesses: manually written and automatically generated. The former involves creating test cases by hand, specifying inputs and expected outcomes. The latter generates diverse test inputs to explore different program states and identify potential issues. Previously  
Automated test generation for Scratch has been explored through various techniques. One early approach models test generation as a search problem using grammatical evolution \cite{deiner2020search}. While effective, it is limited by the slow execution of UI-driven tests and the complexity of instrumenting Scratch’s virtual machine for fitness evaluation. Later, WHISKER was extended with model-based testing, allowing users to define state diagrams from which test cases could be automatically derived \cite{gotz2022model}. However, abstract state definition remains a challenge due to its reliance on domain-specific knowledge, which affects model accuracy and completeness.
Other efforts incorporated neuroevolution with search-based testing to adaptively generate test cases \cite{feldmeier2022neuroevolution}. Although promising, this method incurred high training time for the neural networks that act as input generators. To mitigate this, subsequent work leveraged human gameplay traces and gradient descent to accelerate training \cite{feldmeier2023learning}. To further improve coverage, random and search-based test generation techniques have been employed \cite{deiner2023automated}. However, conventional search techniques still struggle to reach advanced program states, particularly in game like projects.
Afterwards, novelty search was introduced to reward behavioral diversity during test generation, helping avoid deceptive fitness landscapes where traditional methods often stagnate \cite{feldmeier2024combining}. While this approach broadens exploration, it introduces challenges in balancing novelty with goal-directed coverage.
Most recently, the NEATEST framework was extended with many-objective optimization algorithms such as MIO and MOSA \cite{feldmeier2025many}. While their work incorporates branch coverage as an evaluation metric, it does so only for benchmarking purposes, without elaborating on how the metric is computed or analyzed. Their primary focus lies in enhancing the optimization strategy itself. In contrast, this study explicitly targets and evaluates branch coverage  within the context of Scratch. Our work introduces structural refinements for selecting coverage targets using control dependence graphs (CDGs), enabling more precise targeting of branching logic. By focusing on how branch coverage is measured and improved in this domain, this study offers a deeper analysis of the metric’s effectiveness and reliability in test generation for Scratch programs.

\section{Methodology: Achieving Branch Coverage in Scratch}
\label{sec:approach}
This section describes how branch coverage is computed in Scratch programs, the enhancements made to NEATEST, and the fitness function used for optimizing neural networks in test generation. The methodology includes three key steps:

\begin{enumerate}
\itemsep0em 
    \item Identification and classification of control structures in Scratch.
    \item Optimizing NEATEST to prioritize branch coverage.
    \item Designing a fitness function to guide search-based software testing effectively.
\end{enumerate}

\subsection{Classification of Control Structures in Scratch}

To effectively compute branch coverage in Scratch, it is essential to categorize control structures based on their execution behavior. Some control structures explicitly introduce decision points, while others influence execution flow without creating a distinct false branch. Understanding these classifications allows for better test generation, ensuring that all relevant execution paths are exercised.

Scratch control structures can be categorized into three main types:  
\begin{enumerate}
\itemsep0em 
    \item Branching Blocks, which introduce explicit decision-making.
    \item No-False-Branch Blocks, which influence execution without alternative conditions.
    \item Execution-Halting Blocks, which pause or stop execution and indirectly affect control flow.
\end{enumerate}

To systematically identify and classify these structures, we define a function called \texttt{ControlFilter}. \texttt{ControlFilter} analyzes Scratch programs, detects conditional statements, loops, and execution-halting blocks, and categorizes them based on their execution properties. This ensures that the test generation process accounts for all significant control flow elements.

\subsubsection{Branching Blocks}
Branching blocks introduce conditional execution paths in a Scratch program. These blocks either execute conditionally based on a given expression or introduce loops with exit conditions.

\paragraph{Single-Branch Blocks}  
Single-branch blocks have only one execution path and do not provide an alternative path if the condition is false. These blocks execute only when their condition is satisfied; otherwise, execution proceeds to the next statement.
The following Scratch blocks fall under this category:

\begin{itemize}
\itemsep0em 
    \item \textbf{\texttt{control\_if}:} Executes a block of code only if the specified condition evaluates to \textbf{true}. If the condition is false, execution skips to the next statement. Since this block lacks an explicit false branch, a test suite optimized only for statement coverage may never test the false condition, leading to incomplete coverage.

    \item \textbf{\texttt{control\_repeat}, \texttt{control\_repeat\_until}, and \texttt{control\_forever}:} These loops terminate only when specific conditions are met. The \texttt{repeat until} loop introduces a \textbf{true/false decision}, requiring test cases that cover
    execution before the condition is met, and 
    execution after the condition is met (loop exit).

    \item \textbf{\texttt{control\_wait\_until}:} Halts execution until a condition is satisfied. Since execution stalls until the condition is satisfied, testing strategies must ensure cases where the condition is true immediately, and situation where
    the condition remains false for an extended period.
\end{itemize}

\paragraph{Double-Branch Blocks}  
Double-branch blocks explicitly introduce two execution paths—a \textbf{true} path and a \textbf{false} path. To satisfy branch coverage, test cases must be generated to exercise both branches.
The primary double-branch block in Scratch is:
\begin{itemize}
\itemsep0em 
    \item \textbf{\texttt{control\_if\_else}:} Evaluates a condition and executes either the \textbf{true branch} or the \textbf{false branch}. Both execution paths must be tested for full coverage. If only one branch executes, faults in the untested branch remain undetected. 
\end{itemize}



\subsubsection{No-False-Branch Blocks}
Some control structures in Scratch do not introduce an explicit false path but still influence execution flow. These blocks execute unconditionally or pause execution temporarily.
The main no-false-branch blocks in Scratch include:
\begin{itemize}
\itemsep0em 
    \item \textbf{\texttt{control\_forever}:} Executes indefinitely without a terminating condition.
    \item \textbf{\texttt{control\_wait\_until}:} Pauses execution until a condition is met.
\end{itemize}

Some other blocks that pause execution temporarily include \texttt{wait}, \texttt{say}, and \texttt{play sound}. These blocks do not introduce explicit \textbf{true/false decisions}, but they influence execution flow and timing. Testing strategies targeting these blocks must account for scenarios where:
\begin{itemize}
\itemsep0em 
    \item The loop runs indefinitely, preventing further execution.
    \item Execution is paused for a delay, affecting test scheduling.
\end{itemize}

\subsubsection{Execution-Halting Blocks}
Execution-halting blocks pause or terminate execution, implicitly affecting program flow. While they do not create explicit branching conditions, they may prevent alternate execution paths from being reached.

The main execution-halting blocks include:
\begin{itemize}
\itemsep0em 
    \item \textbf{\texttt{control\_stop}:} Immediately terminates execution.
    \item \textbf{\texttt{control\_wait}:} Delays execution for a set duration.
\end{itemize}

If a \textbf{stop block} is reached in a true branch, the false branch may remain untested. Test cases must ensure that:
\begin{itemize}
\itemsep0em 
    \item Execution is forced beyond the stop block.
    \item Alternate paths that avoid stopping the script are explored.
\end{itemize}

\subsubsection{Role of ControlFilter in Test Generation}
To ensure comprehensive branch coverage, we employ \texttt{ControlFilter} to systematically identify and classify control structures. \texttt{ControlFilter} first detects all control blocks within the program by filtering opcodes that start with \texttt{control\_}. It then categorizes them into branching blocks, no-false-branch blocks, and execution-halting blocks. By distinguishing between these types, \texttt{ControlFilter} enables the test generator to target conditional paths more effectively, ensuring that all relevant execution states are evaluated.

\subsection{Neatest with Statement Coverage ($N_{SC}$)}
\label{ssec:nsc}
NEATEST is a neuroevolution-based test suite generator designed for Scratch games. It evolves neural networks as test agents, optimizing them to interact with a Scratch program and generate test cases that maximize coverage. By default, NEATEST optimizes statement coverage, ensuring that every statement in the program is executed at least once \cite{feldmeier2022neuroevolution}.


Neatest with Statement Coverage ($N_{SC}$) is a test generation strategy designed to maximize statement coverage in Scratch programs. By evolving neural networks as test agents, $N_{SC}$ ensures that each executable statement in a Scratch program is covered at least once. However, while this approach is effective in reaching individual program statements, it does not explicitly account for the execution of different branching conditions, leading to potential gaps in the evaluation of program logic.

$N_{SC}$ operates by selecting uncovered statements as test targets. The process begins with the identification of the next uncovered statement using the Control Dependence Graph (CDG), after which a population of neural networks evolves towards executing that statement. Each network is assigned a fitness score based on its effectiveness in reaching the desired statement, and once coverage is achieved, the test generator moves on to the next uncovered statement. This process continues iteratively until all statements in the program have been executed at least once. While this approach provides a measure of completeness, it does not ensure that the logical decisions governing program flow are thoroughly tested.

The primary limitation of $N_{SC}$ lies in its inability to guarantee branch coverage, particularly in cases where decision structures introduce alternative execution paths. Consider the Scratch program shown in Figure~\ref{fig:scratch-code}, which consists of a variable initialization, a conditional check for key input, a conditional assignment, and a movement command. If a test case generated by $N_{SC}$ executes the if-condition where the space key is pressed, the statement inside the condition (\texttt{set my variable to 0}) will be covered, along with the subsequent movement command (\texttt{move my variable steps}). However, $N_{SC}$ does not ensure that the alternative execution path, where the space key is not pressed, is ever tested. In this scenario, if the variable remains \texttt{null}, executing the move command might lead to an error or unexpected behavior. Since $N_{SC}$ does not systematically generate test cases that force the program into both possible execution paths, such issues may go undetected.

This limitation becomes critical when testing programs with multiple branching conditions, as $N_{SC}$ may inadvertently favor one execution path while neglecting others. This can result in a false sense of test completeness, where all statements have been executed, yet critical faults in untested branches remain hidden. The inability to enforce execution of all decision-making constructs makes $N_{SC}$ insufficient for rigorous testing, particularly in scenarios where program correctness depends on the behavior of conditional logic.

To address these shortcomings, Neatest with Branch Coverage ($N_{BC}$) extends $N_{SC}$ by explicitly prioritizing branch coverage as an optimization goal. Unlike $N_{SC}$, which merely ensures execution of individual statements, $N_{BC}$ guarantees that both the true and false branches of every conditional statement are exercised. This enhancement not only improves test completeness but also strengthens fault detection by ensuring that all decision-based execution paths in a Scratch program are systematically explored. The following section describes how $N_{BC}$ overcomes the limitations of $N_{SC}$ by integrating a structured approach for prioritizing uncovered branches during test generation.

\subsection{Neatest with Branch Coverage ($N_{BC}$)}
\label{ssec:nbc}

Neatest with Branch Coverage ($N_{BC}$) extends $N_{SC}$ by ensuring that both true and false branches of conditional statements are executed. Unlike $N_{SC}$, which focuses solely on executing statements, $N_{BC}$ explicitly targets unexecuted branches to provide a more comprehensive evaluation of program logic. This improvement allows $N_{BC}$ to identify faults that $N_{SC}$ may overlook, particularly those related to unexplored execution paths in decision-making structures.

$N_{BC}$ modifies NEATEST’s optimization strategy by prioritizing branch execution over statement execution. Instead of selecting uncovered statements as test targets, $N_{BC}$ selects uncovered branches using the Control Dependence Graph (CDG). The core process of $N_{BC}$ is detailed in Algorithm~\ref{algo:nbc}, which dynamically selects branches, evolves test agents using neuroevolution, and validates their robustness before including them in the final test suite.

\subsubsection{Overview}
The Neatest with Branch Coverage ($N_{BC}$) algorithm (Algorithm \ref{algo:nbc}) extends the Neatest with Statement Coverage ($N_{SC}$) approach by systematically ensuring that both true and false branches of conditional statements are executed during test generation. Unlike $N_{SC}$, which focuses on reaching uncovered statements, $N_{BC}$ dynamically prioritizes unexecuted branches as test targets to improve logical path exploration and fault detection.

The algorithm begins by initializing a requireNextBranch flag (line 2) to indicate whether a new target branch should be selected. The execution proceeds in an iterative loop (line 3), terminating only when the stopping condition is met (e.g., reaching a time limit or achieving full branch coverage). If a new target branch is required (line 4), the \texttt{selectTargetBranch()} function (line 5) is called. This function queries the Control Dependence Graph (CDG) to identify an unexecuted branch that should be selected for testing. The ControlFilter module further ensures that control structures are correctly categorized and analyzed to facilitate optimal branch selection.

Once the target branch is determined, a population of neural networks is generated (line 6) using \texttt{generate}-\texttt{Population()}. These networks serve as candidate solutions, evolving through neuroevolution to trigger the selected branch. The algorithm then enters a training phase, where each network executes the program (line 9) and computes a fitness score (line 10) based on its success in executing the target branch.

If a network successfully covers the target branch (line 11), it undergoes a robustness check (line 12) via the \texttt{robustnessCheck()} function. This step ensures that the network can reliably execute the same branch across multiple program executions under different conditions. The network's fitness is updated accordingly (line 13), and if it meets the desired robustness count (line 15), it is added to the dynamic test suite (line 16). The branch is then marked as covered in $B_u$ (line 17), and the algorithm sets the flag to requireNextBranch = true (line 18), triggering selection of the next uncovered branch.

If no network has successfully covered the branch, the NEAT evolutionary process (line 21) is triggered to mutate and refine the population for improved performance in subsequent iterations. This cycle repeats until all branches are covered or the search budget is exhausted.

The branch selection process (line 5) is further detailed in Section \ref{sssec:branch-selection-strategy}, where the selection strategy ensures that execution history and structural dependencies are considered when selecting the next target branch. This strategic prioritization allows $N_{BC}$ to intelligently shift focus between branches, ensuring comprehensive branch coverage while minimizing redundant test executions.

By integrating ControlFilter and refining the branch selection strategy, $N_{BC}$ achieves a more structured and efficient test evolution compared to $N_{SC}$. The proceeding section details the fitness function (Section \ref{sssec:fitness}), which quantifies how well a test case triggers a given branch and guides neuroevolution toward optimal solutions.

\begin{algorithm}[t]

\caption{Neatest with Branch Coverage ($N_{BC}$)}
\label{algo:nbc}

    \includegraphics[width=0.5\linewidth]{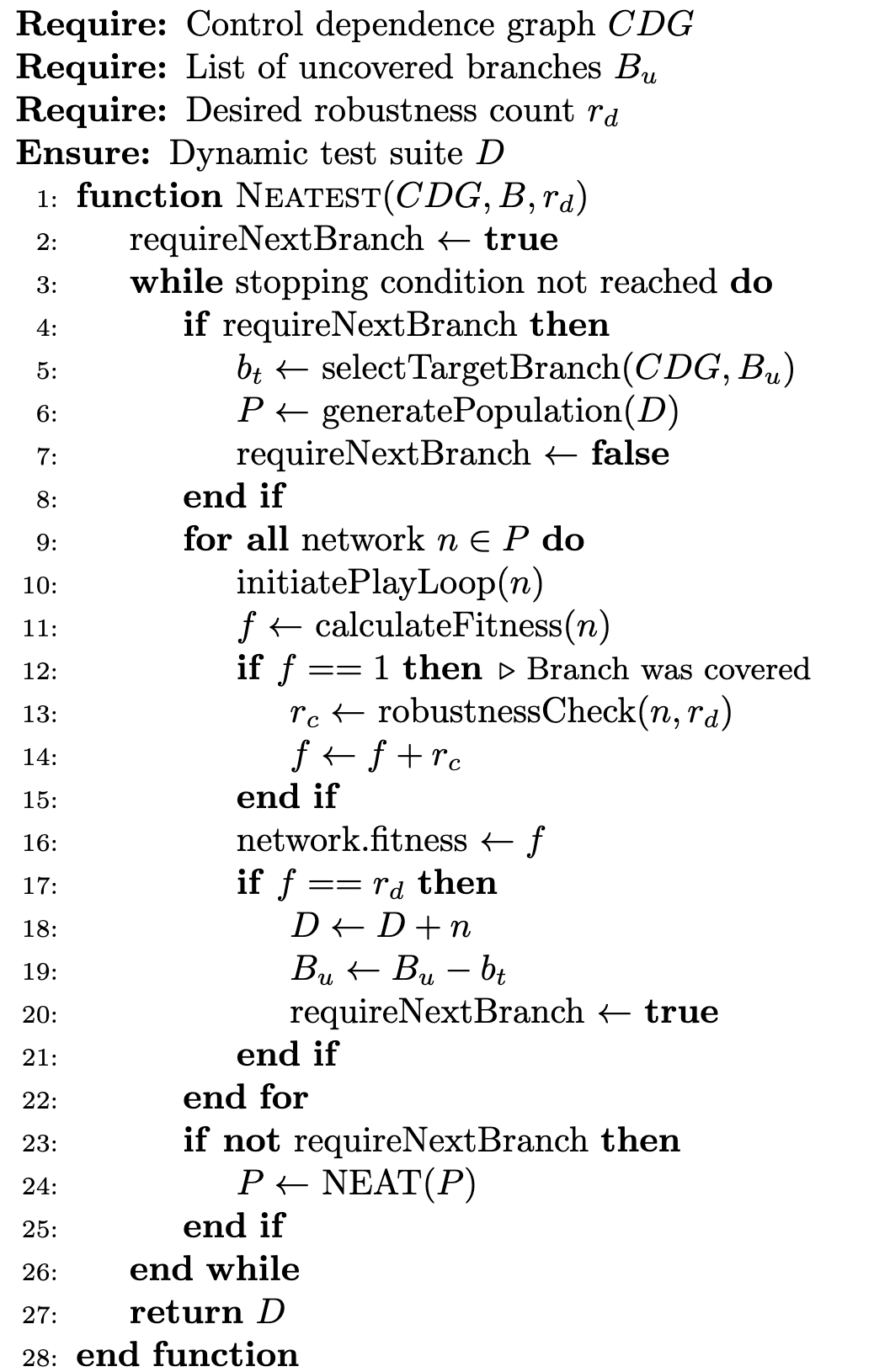}
\end{algorithm}

\subsubsection{Branch Selection Strategy}
\label{sssec:branch-selection-strategy}
A crucial step in $N_{BC}$ is the selection of the next uncovered branch to be tested (line 5 in Algorithm~\ref{algo:nbc}). Instead of selecting branches randomly, $N_{BC}$ employs a guided selection approach using ControlFilter, which continuously analyzes the Control Dependence Graph (CDG) to determine which branches remain unexecuted.

ControlFilter plays a pivotal role in ensuring that test generation is both systematic and efficient. Its primary functions include:

\begin{enumerate}
\itemsep0em 
    \item \textbf{Filtering Out Non-Branching Structures:} Scratch programs contain loops and control blocks that do not contribute to branch coverage. ControlFilter ensures that the test generation process focuses only on relevant conditional branches.
    \item \textbf{Tracking Uncovered Branches:} ControlFilter maintains a list of branches that have not been executed yet, ensuring that these branches are targeted before redundant executions.
    \item \textbf{Guiding Selection Strategy:} Rather than selecting branches arbitrarily, ControlFilter prioritizes execution paths based on two key factors:
    \begin{itemize}
    \itemsep0em 
        \item \textbf{Execution History:} If a branch has already been executed at least once, $N_{BC}$ shifts focus to an unexecuted branch within the program. This prevents redundant test executions and ensures systematic exploration of new logic paths.
        \item \textbf{Structural Dependencies:} If a branch leads to deeper, unexplored code segments, it is prioritized over branches that lead to already-covered regions. This ensures that $N_{BC}$ progressively explores the program structure rather than stagnating on easy-to-reach branches.
\end{itemize}
\end{enumerate}

For example, in a Scratch game where a character collects different items (e.g., coins and apples), the initial test case might focus on collecting coins. If, during evolution, the same test incidentally collects apples, $N_{BC}$ then prioritizes testing for apple collection in future iterations. By doing so, $N_{BC}$ ensures that all branches are executed intentionally, rather than relying on incidental test coverage.

ControlFilter enhances branch selection by refining the search space, directing the test generator toward meaningful execution paths, and reducing wasted test efforts. Without ControlFilter, $N_{BC}$ could end up exploring execution paths inefficiently, leading to longer optimization times and incomplete branch coverage.

\subsubsection{Fitness Evaluation}
\label{sssec:fitness}
In Neatest with Branch Coverage ($N_{BC}$), neural networks evolve to maximize their ability to execute both branches of conditional statements. To guide this evolution, each network is evaluated using a fitness function that quantifies its effectiveness in reaching and executing an uncovered branch. The fitness score is based on two fundamental factors:

\begin{enumerate}
\itemsep0em 
    \item \textbf{Approach Level} ($\alpha$) – Represents how far execution is from reaching the control dependency of the target branch.
    \item \textbf{Branch Distance} ($\beta$) – Measures how much execution deviates from taking the desired branch once the control statement is reached.
\end{enumerate}
The fitness function integrates these two metrics to prioritize networks that are closest to executing an untested branch. Networks with lower approach levels and smaller branch distances receive higher fitness scores, increasing their likelihood of being selected for further evolution.
Since both \(\alpha\) and \(\beta\) can take on large values, they are normalized to ensure fair contribution to the fitness score. We define the normalization function as follows:

\begin{equation}
    \eta(x) = \frac{x}{1 + x}
\end{equation}

This function ensures that all values are scaled into the range \( (0,1] \), preventing excessive influence from large values.
The fitness score for a neural network is computed using the normalized approach level and branch distance:

\begin{equation}
\label{eq:fitness}
    f = \eta \Big(\alpha + \eta(\beta) \Big)
\end{equation}

\begin{itemize}
\itemsep0em
    \item \( \alpha \) (Approach Level): Determines the number of control dependencies between the execution point and the target branch. If  \(\alpha=0\), the execution has already reached the control statement. 
    \item \( \beta \) (Branch Distance): Represents how close the execution is to taking the desired branch. A smaller \( \beta \) means the network is closer to executing the correct branch and \( \beta = 0 \) means that the branch has been executed. 
\end{itemize}

If the fitness function results in 
$f=0$, it means that the target branch has been successfully executed. At this point, robustness testing is performed to verify whether the network can reliably execute the same branch across multiple randomized test runs. If the network passes robustness testing, it is added to the final dynamic test suite, ensuring stable test generation.

\begin{lstlisting}[style=pseudoStyle, caption={Code Sample 2.}, label={lst:cfg}]
When green flag clicked
    If (X > 0)
        If (Y = 0)
            Say "C"
        Else
            Say "B"
    Else
        Say "A"
\end{lstlisting}
\paragraph*{\textbf{Computing Fitness Score: Example Walkthrough}\\ }
To effectively evolve test cases that maximize branch coverage, $N_{BC}$ relies on branch distance as a key component of its fitness function. Branch distance quantifies how far an execution path is from taking the desired branch in a conditional statement. The smaller the branch distance, the closer the neural network is to triggering the correct execution path.

Branch distance is computed differently depending on whether the approach level is satisfied. If the approach level is greater than zero, meaning the execution has not yet reached the control dependency of the target branch, then branch distance is calculated for the nearest control dependency rather than the branch itself. If the approach level is zero, meaning the execution has reached the target control statement, branch distance measures how close the execution is to taking the correct branch.

\begin{figure}[h]  
    \centering

    \begin{subfigure}{0.45\linewidth}  
        \centering
        \includegraphics[width=\linewidth]{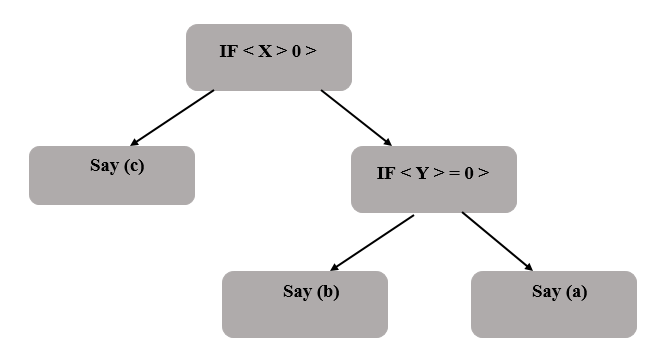}
        \caption{CFG}
        \label{fig:cfg}
    \end{subfigure}
    \hfill
    \begin{subfigure}{0.45\linewidth}  
        \centering
        \includegraphics[width=\linewidth]{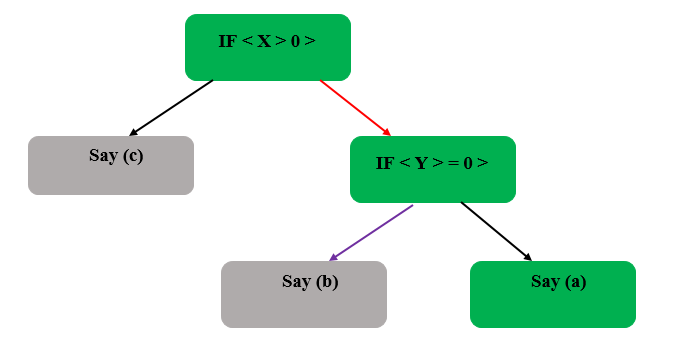}
        \caption{Visualization of node execution when approach level is zero.}
        \label{fig:cfg1}
    \end{subfigure}
    \hfill
    \begin{subfigure}{0.5\linewidth}  
        \centering
        \includegraphics[width=\linewidth]{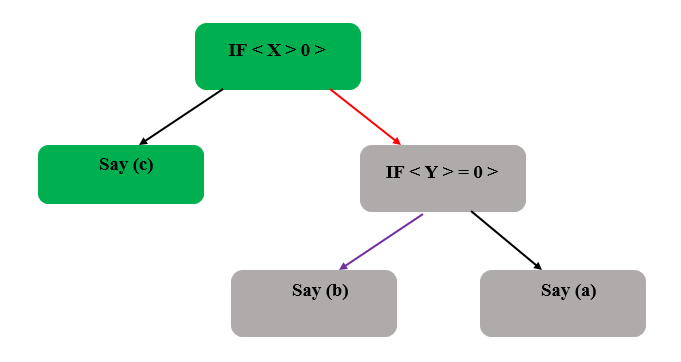}
        \caption{Visualization of node execution when approach level is greater than zero.}
        \label{fig:cfg2}
    \end{subfigure}

    \caption{CFG corresponding to Listing \ref{lst:cfg}.}
    \label{fig:multi-cfg}
\end{figure}

Consider the Scratch program represented in Listing \ref{lst:cfg} and its corresponding CFG shown in Figure \ref{fig:cfg}. This program consists of nested conditional statements where the execution follows different branches based on the values of \texttt{X} and \texttt{Y}. The objective is to compute the branch distance required to execute a specific branch.


\paragraph*{\textbf{Case 1: Approach Level  = 0}: } 

Suppose the goal is to execute the false branch of \texttt{If (Y = 0)} and the input values are \texttt{X = 10, Y = 1}. As the execution reaches the condition \texttt{If (Y = 0)}, it evaluates to false, leading to the execution of the true branch (\texttt{Say (B)}) instead of the target false branch (\texttt{Say (C)}) as shown on Figure~\ref{fig:cfg1} . 

The branch distance in this case is calculated as:

\[
\beta = |Y - 0| = |1 - 0| = 1
\]

This indicates that reducing \texttt{Y} to zero would have resulted in correct execution.

\paragraph*{\textbf{Case 2: Approach Level $>$ 0}: }
Consider an alternative input where \texttt{X = -10, Y = -1}. Since \texttt{X <= 0}, execution follows the false branch of \texttt{If (X > 0)}, leading to \texttt{Say (A)} and bypassing \texttt{If (Y = 0)} entirely as shown in Figure~\ref{fig:cfg2}.  

In this case, the branch distance is calculated relative to the closest control dependency.

Since execution never reached the condition \texttt{If (Y = 0)}, the approach level is determined first. The execution must first satisfy \texttt{If (X > 0)} before it can reach \texttt{If (Y = 0)}, making the approach level:

\[
\alpha = 1
\]

The branch distance for the unmet condition \texttt{If (X > 0)} is calculated as:

\[
\beta = |X - 0| = |-10 - 0| = 10
\]

This indicates that increasing \texttt{X} to be greater than zero would allow execution to progress toward the next conditional branch.

\paragraph*{\textbf{Fitness Calculation}:}
Once the approach level and branch distance are computed, the fitness function (Equation~\ref{eq:fitness}) is applied.

For \textbf{Case 1} (\texttt{X=10, Y=1}), where approach level is zero and branch distance is 1, fitness is calculated as follows:

\[
f = \eta \Big(0 + \eta(1)\Big) = \eta \Big(0 + \frac{1}{1+1}\Big) = \eta \Big(0 + 0.5\Big) =  \frac{1}{1+0.5} = 0.666
\]


For \textbf{Case 2} (\texttt{X=-10, Y=-1}),
where the approach level is one and the branch distance is ten, fitness is calculated as follows:

\[
f = \eta\Big(1 + \eta(10)\Big) = \eta\Big(1 + \frac{10}{1+10}\Big) = \eta\Big(1.91\Big)=\frac{1.91}{1+1.91}= 0.656
\]

The computed fitness score directly influences the evolution of neural networks in $N_{BC}$. Higher fitness scores indicate that test cases need further improvement. Through mutation and crossover, networks with lower fitness values (closer to zero) are favored in the selection process, driving the neuroevolution towards executing all untested branches.
This iterative approach ensures that all logical paths are covered, thereby improving both branch coverage and fault detection capabilities in Scratch programs.

\section{Evaluation Setup}
\label{sec:evaluation}

This section describes the experimental setup used to evaluate the effectiveness of Neatest when optimized for branch coverage ($N_{BC}$) compared to statement coverage ($N_{SC}$). The evaluation is structured around two core research questions (RQs), each designed to measure different aspects of test generation performance. To ensure comprehensive and reproducible results, we conduct experiments using a diverse set of Scratch programs, employ statistical significance tests, and analyze the impact of branch-aware test generation on both test coverage and fault detection capabilities.

\subsection{Research Questions}

To evaluate the effectiveness of the proposed testing approach, we aim to answer the following research questions:

\begin{enumerate}[label=\textbf{RQ\arabic*}]

    \item \textbf{How does test coverage differ when NEATEST is optimized for branch coverage ($N_{BC}$) compared to statement coverage ($N_{SC}$)?}\\ Here, test coverage refers to both statement coverage (SC) and branch coverage (BC). Since $N_{SC}$ is designed to maximize statement coverage, it may achieve partial branch coverage as a side effect. However, $N_{BC}$ explicitly targets branch coverage, ensuring that both true and false branches of conditional statements are executed. This research question examines whether $N_{BC}$ achieves superior branch coverage while maintaining comparable statement coverage to $N_{SC}$.
    \item \textbf{How does test oracle effectiveness vary when NEATEST is optimized for branch coverage ($N_{BC}$) compared to statement coverage ($N_{SC}$)?}\\ While high coverage is crucial, it does not always correlate with the ability to detect program faults. To assess test oracle effectiveness, we measure the mutation score, which represents the percentage of artificially injected faults (mutants) that the generated tests can detect. If $N_{BC}$ provides better branch coverage, we expect it to generate more effective test cases capable of revealing faults in conditional logic.
\end{enumerate}

NEATEST is integrated as part of the Whisker framework and is publicly available as an open-source tool\footnote{[August 2022] https://github.com/se2p/whisker
}. The dataset for our experiments, raw experimental results, and reproduction scripts are also available on GitHub.

\subsection{Dataset}
\label{sec4.1}

We reuse the dataset established in the NEATEST study \cite{feldmeier2022neuroevolution} for evaluation. The dataset comprises 187 Scratch programs collected from beginner's Scratch programming books\footnote{[August 2022] https://nostarch.com/catalog/scratch}, prior studies, and online tutorial websites (Code Club\footnote{[August 2022] https://projects.raspberrypi.org/en/codeclub}, Linz Coder 
Dojo\footnote{[August 2022] https://coderdojo-linz.github.io/uebungsanleitungen/ programmieren/scratch/}, learnlearn\footnote{[August 2022] https://learnlearn.uk/scratch/}, junilearning\footnote{[August 2022] https://junilearning.com/blog/coding-projects/})
. The dataset was curated based on three selection criteria:

\begin{enumerate}
\itemsep-0.2em
    \item The program must be a game that accepts user inputs, processes them through the Whisker framework, and presents a challenge to the player.
    \item The game must exhibit dynamic behavior, requiring reactive test case generation.
    \item Programs that lack a meaningful challenge—where arbitrary input could trivially achieve full coverage—were excluded.
\end{enumerate}

After filtering based on these conditions, the final dataset comprises 25 unique Scratch games, each exhibiting distinct interaction challenges. These games are presented in Table~\ref{tab:evaluation_games}.

\begin{table*}[t]
    \centering
    \scriptsize
    \caption{Overview of the 25 Scratch game projects used for evaluation, detailing the number of sprites, scripts, and executable statements in each.}
    \label{tab:evaluation_games}
    \begin{tabular}{|l|ccc|l|ccc|}
        \hline
         \textbf{Projects} & \rotatebox{90}{\textbf{\#Sprites}} & \rotatebox{90}{\textbf{\#Scripts}} & \rotatebox{90}{\textbf{\#Statements}} & \textbf{Projects} & \rotatebox{90}{\textbf{\#Sprites}} & \rotatebox{90}{\textbf{\#Scripts}} & \rotatebox{90}{\textbf{\#Statements}} \\
        \hline
         FlappyParrot & 2 & 7 & 37 & FinalFight & 13 & 48 & 286 \\
        Frogger & 8 & 22 & 107 & OceanCleanup & 11 & 22 & 156 \\
         Fruitcatching & 3 & 4 & 55 & Pong & 2 & 2 & 15 \\
         LineUp & 2 & 4 & 50 & RioShootout & 8 & 26 & 125 \\
        BrainGame & 3 & 18 & 76 & Dragons & 6 & 33 & 381 \\
         EndlessRunner & 8 & 29 & 163 & WhackAMole & 10 & 14 & 391 \\
         Snake & 3 & 14 & 60 & SpaceOdyssey & 4 & 13 & 116 \\
         CatchTheDots & 4 & 10 & 82 & SnowballFight & 3 & 6 & 39 \\
        CatchTheGifts & 3 & 7 & 68 & HackAttack & 6 & 9 & 93 \\
         CityDefender & 10 & 12 & 97 & DieZauberlehrlinge & 4 & 14 & 87 \\
         CatchingApples & 2 & 3 & 25 & Dodgeball & 4 & 10 & 78 \\
         DesertRally & 10 & 27 & 212 & BirdsShooter & 4 & 10 & 69 \\ \cline{5-8}
         FallingStars & 4 & 4 & 91 & Mean & 5.5 & 16.6 & 118.3 \\
        \hline
    \end{tabular}
\end{table*}
\normalsize
\subsection{Experimental Setup}
\label{sec4.2}

All experiments were conducted on a dedicated computing cluster consisting of nine nodes, each powered by an AMD EPYC 7443P CPU running at 2.8 GHz. The Whisker framework allows for an acceleration factor to speed up test execution. To ensure efficient testing, we set this factor to infinity (\(\infty\)), maximizing execution speed.

\subsubsection{Configuration for RQ1: Coverage Comparison Between Statement and Branch Coverage}

To investigate how test coverage varies when NEATEST is optimized for branch coverage, we extended the framework by introducing a fitness function specifically designed to target branch coverage. This was compared against the existing statement coverage function, which, while focused on maximizing statement coverage, may incidentally cover branches without explicitly optimizing for them. 
Each experiment was conducted under the following conditions:
\begin{itemize}
\itemsep-0.2em
    \item Population size of 150 neural networks for each fitness function.
    \item Search duration of 5 hours.
    \item Non-improving generation limit set to 5.
    \item Species size of 5.
    \item Maximum playthrough duration of 5 seconds (equivalent to 100 seconds due to acceleration).
\end{itemize}

 To ensure statistical robustness, we repeated the experiment 30 times ( $\mathcal{N} = 30$), following the recommendations of Arcuri et al. \cite{arcuri2011practical}.
Independent repetitions were performed for each fitness function \cite{arcuri2011practical}. The Vargha and Delaney (A12) effect size \cite{vargha2000critique} and the Mann-Whitney U-test \cite{mann1947test} (\(\alpha = 0.05\)) were applied to assess statistical significance.

\subsubsection{Configuration for RQ2: Mutation Analysis for Test Oracle Effectiveness}
To assess whether neural networks can function effectively as test oracles, we conducted a mutation analysis using eight predefined mutation operators (Table~\ref{tab:mutation_operators}) drawn from a widely accepted set of sufficient mutation operators \cite{offutt1996experimental}. These operators were applied to all 25 selected Scratch games. A mutant was classified as ``killed" under either of the following conditions: (1) the activation values of the neural network exceeded a predefined threshold of 30, based on Likelihood-based Surprise Adequacy (LSA) \cite{feldmeier2022neuroevolution}, or (2) the network's structure underwent a significant change during execution.

For activation-based evaluation, we employed LSA \cite{kim2018guiding}, which leverages Kernel Density Estimation (KDE) \cite{rosenblatt1956remarks} to assess how unexpected or ``surprising" the activations of selected neurons are when the network is presented with mutated inputs.

To ensure robustness and mitigate the influence of randomness inherent to game behavior, each extracted test suite was evaluated across ten different random seeds, resulting in 100 executions per test suite per game. While RQ1 used 30 repetitions to assess coverage, the increased behavioral variability in mutation analysis justifies a higher repetition count. Prior studies have found 100 runs to provide a reliable basis for mutation-based evaluation and to minimize noise from incidental execution shifts \cite{feldmeier2022neuroevolution}.
Additionally, to manage the risk of combinatorial explosion in the number of mutant variants, each experiment was constrained to 50 randomly selected mutants per mutation operator per repetition.



\begin{table}[t]
    \centering
    \footnotesize
    \caption{Overview of the eight mutation operators applied during evaluation. These operators introduce syntactic changes to Scratch programs to simulate faults and assess the fault detection capability of test oracles, as defined in \cite{feldmeier2022neuroevolution}.}
    \label{tab:mutation_operators}
    \scriptsize
    \begin{tabular}{|l|l|}
        \hline
        \textbf{Mutation Operator} & \textbf{Description} \\
        \hline
        Key Replacement Mutation & Replaces a block key’s listener. \\
        \hline
        Single Block Deletion & Removes a single block. \\
        \hline
        Script Deletion Mutation & Deletes all blocks of a given script. \\
        \hline
        Arithmetic Operator Replacement & Replaces an arithmetic operator. \\
        \hline
        Logical Operator Replacement & Replaces a logical operator. \\
        \hline
        Relational Operator Replacement & Replaces a relational operator. \\
        \hline
        Negate Conditional Mutation & Negates Boolean blocks. \\
        \hline
        Variable Replacement Mutation & Replaces a variable. \\
        \hline
    \end{tabular}
\end{table}




\subsection{Evaluation Metrics}
In this section, we present the definition of the evaluation metrics: statement coverage and decision coverage (RQ1), and, mutation score (RQ2).

For \textit{RQ1}, we use test coverage in terms of statement coverage percentage (SC) as depicted in equation \ref{eq:codecoverage} and branch coverage percentage (BC) as depicted in equation \ref{eq:decisioncoverage} . 

Additionally, for \textit{RQ2}, We consider the mutation score (as shown in equation \ref{eq:mutationscore}) as a criterion to discover whether the test cases have a good bug-revealing ability. 
It is calculated as the ratio of the number of killed mutants to the total number of mutants.

\begin{equation}\label{eq:mutationscore}
\text{Mutation Score} = \frac{\text{Killed Mutants}}{\text{Total Mutants}}
\end{equation}

\section{Results and Analysis}
\label{sec:results}

\subsection{RQ1: Test Coverage}
\label{ssec:rq1}

To assess the effectiveness of Neatest with Branch Coverage ($N_{BC}$) compared to Neatest with Statement Coverage ($N_{SC}$), we conducted experiments on 25 games in our dataset. The primary objective was to determine whether $N_{BC}$ achieves higher test coverage ratios than $N_{SC}$.

\subsubsection{Branch Coverage}
The average branch coverage percentages obtained for $N_{SC}$ and $N_{BC}$ across 30 experimental runs for each game are summarized in Table ~\ref{tab:branch_coverage} 
and shown in figure \ref{fig:cov}. The results indicate that $N_{SC}$ achieved a mean branch coverage of 84.30\%, whereas $N_{BC}$ achieved 86.31\%, demonstrating a slight improvement. However, deeper examination reveals interesting trends based on the structural characteristics of individual games.

\begin{table}[h]
    \centering
    
      \caption{Mean branch coverage\% for \(N_{SC}\) and \(N_{BC}\) for each case study for 30 independent runs along with p-values showing significance and effect size}
    \label{tab:branch_coverage}
    
    \centering
      \scriptsize
    \begin{tabular}{|c|l|c|c|c|c|}
        \hline
   \rowcolor{white} \# & Project & $N_{SC}$ & $N_{BC}$ & P & A12 \\
        \hline
        \rowcolor{yellow!70} 1 & LineUp & 90.47 & 100.00 & 0.00 & 0.03 \\
        \rowcolor{yellow!70} 2 & BrainGame & 98.12 & 100.00 & 0.00 & 0.35 \\
        \rowcolor{yellow!70} 3 & EndlessRunner & 59.26 & 66.66 & 0.00 & 0.24 \\
        \rowcolor{yellow!70} 4 & CatchTheGifts & 96.94 & 100.00 & 0.00 & 0.18 \\
        \rowcolor{yellow!70} 5 & DesertRally & 95.25 & 95.86 & 0.00 & 0.17 \\
        \rowcolor{yellow!70} 6 & OceanCleanup & 72.04 & 75.64 & 0.01 & 0.31 \\
        \rowcolor{yellow!70} 7 & WhackAMole & 73.42 & 77.77 & 0.00 & 0.28 \\
        \rowcolor{yellow!70} 8 & SpaceOdyssey & 90.49 & 91.21 & 0.01 & 0.33 \\
        \rowcolor{yellow!70} 9 & SnowballFight & 75.94 & 83.04 & 0.00 & 0.09 \\
        \rowcolor{yellow!70} 10 & HackAttack & 95.52 & 97.63 & 0.00 & 0.28 \\
        \rowcolor{yellow!70} 11 & DieZauberlehrlinge & 36.77 & 45.16 & 0.01 & 0.31 \\
        \rowcolor{green!30} 12 & Froger & 91.33 & 90.67 & 0.02 & 0.64 \\  
        \rowcolor{green!30} 13 & Dodgeball & 91.08 & 89.68 & 0.02 & 0.665 \\
        \rowcolor{orange!50} 14 & FlappyParrot & 81.21 & 82.12 & 0.08 & 0.45 \\
        \rowcolor{orange!50} 15 & FruitCatching & 90.71 & 91.78 & 0.08 & 0.37 \\
        \rowcolor{orange!50} 16 & Snake & 77.85 & 78.80 & 0.15 & 0.43 \\
        \rowcolor{orange!50} 17 & CityDefender & 63.87 & 64.41 & 0.18 & 0.45 \\
        \rowcolor{orange!50} 18 & Dragons & 49.66 & 52.75 & 0.11 & 0.38 \\
        \rowcolor{gray!50} 19 & CatchingApples & 100.00 & 100.00 & 1.00 & 0.5 \\
        \rowcolor{gray!50} 20 & BirdsShooter & 100.00 & 100.00 & 1.00 & 0.5 \\
        \rowcolor{gray!50} 21 & RioShootout & 94.54 & 94.54 & 1.00 & 0.5 \\
        \rowcolor{red!50} 22 & CatchTheDots & 90.76 & 89.14 & 0.06 & 0.60 \\
        \rowcolor{red!50} 23 & Pong & 100.00 & 98.88 & 0.16 & 0.53 \\
        \rowcolor{red!50} 24 & FallingStars & 94.16 & 93.78 & 0.23 & 0.57 \\
        \rowcolor{red!50} 25 & FinalFight & 98.20 & 98.15 & 0.34 & 0.43 \\
        \hline
        \rowcolor{white} & Mean & 84.30 & 86.31 & - & 0.3836 \\

 \hline
    \end{tabular}
  
\end{table}

\paragraph*{Case 1: $N_{BC}$ Significantly Outperforms $N_{SC}$ \\ }

$N_{BC}$ outperforms $N_{SC}$ significantly in 13 games, with statistically significant p-values, indicating that $N_{BC}$ consistently achieves higher branch coverage (as shown in Table \ref{tab:branch_coverage}). These games include:
BrainGame, LineUp, EndlessRunner, HackAttack,
CatchTheGifts, DesertRally, OceanCleanup, WhackAMole, SpaceOdyssey, SnowballFight, and DieZauberlehrlinge.

The improved performance of $N_{BC}$ in these games suggests that its ability to force execution of both true and false branches allows for greater structural exploration, resulting in higher coverage ratios.

\paragraph*{Case 2: $N_{BC}$ Achieves Higher Mean Coverage, but Not with Statistical Significance \\ }

In 5 games, $N_{BC}$ achieved better mean coverage than $N_{SC}$, but the p-values exceeded 0.05 (as shown in Table \ref{tab:branch_coverage}), indicating high variance in performance across runs. These games are:
FlappyParrot, FruitCatching, Snake, CityDefender, and Dragons.

The inconsistency in $N_{BC}$’s performance can be attributed to variability in evolved test strategies. Additionally, static analysis of the game structures reveals that several if-conditions are mutually exclusive, meaning that achieving one condition automatically negates the other. This reduces the advantage of explicitly optimizing for branch coverage. 

For example, in FlappyParrot and Snake, the primary control structures are forever loops and halting blocks, which do not significantly impact branch coverage.
In FruitCatching and CityDefender, the most valuable sprites for branch coverage contain if-conditions that are mutually exclusive. In FruitCatching, the bowl sprite has two if-conditions—one to move left, the other to move right. Similarly, the apple sprite has two if-conditions—one checking if it touches the bowl, the other checking if it touches the ground.
In CityDefender, the meteorite sprite has two if-conditions—one checking for impact with the city, the other for impact with a color-coded target.
In these cases, branch coverage behaves similarly to statement coverage, reducing $N_{BC}$’s advantage. In the Dragons game, $N_{BC}$’s search process requires more sophisticated evolution strategies to optimize coverage.

\paragraph*{Case 3: $N_{BC}$ and $N_{SC}$ Achieve Equal Coverage \\ }

In 3 games, $N_{SC}$ and $N_{BC}$ achieved nearly identical branch coverage percentages (as shown in Table \ref{tab:branch_coverage}), making them functionally equivalent. These games are:
CatchingApples, BirdsShooter, and RioShootout.

Static analysis of these games reveals that they have very few control dependencies (i.e., few branch conditions). Hence, the allocated search budget was sufficient for both methods to achieve full coverage.
In RioShootout, even though the number of statements and sprites is higher, the key sprite (ball) has two independent if-conditions—one for scoring a goal and one for tracking missed goals. These if-conditions are mutually exclusive, meaning that statement coverage automatically results in branch coverage.

\paragraph*{Case 4: $N_{SC}$ Achieves Higher Mean Coverage, but Not with Statistical Significance
\\ }

In 4 games, $N_{SC}$ achieved slightly higher mean coverage than $N_{BC}$. These games are:
CatchTheDots, Pong, FallingStars, and FinalFight.

Although the differences in coverage were small (0-2\%), this suggests that $N_{SC}$ performs better in games with simpler control dependencies or higher randomness.

Pong has the fewest number of statements and sprites, meaning that full coverage can be achieved efficiently using $N_{SC}$.
FinalFight contains a high number of statements but very few control structures, reducing the difference between $N_{SC}$ and $N_{BC}$’s performance.
Additionally, intrinsic randomness in games like FallingStars means that test generators using $N_{SC}$ may sometimes incidentally trigger more if-conditions than $N_{BC}$.

\paragraph*{Case 5: $N_{SC}$ Significantly Outperforms $N_{BC}$ \\ }
In Froger and Dodgeball, $N_{SC}$ attains a slightly higher mean branch coverage than $N_{BC}$ (as shown in Table \ref{tab:branch_coverage}). The p-values (0.02) indicate that $N_{SC}$'s performance is statistically significant, meaning the observed difference is unlikely due to chance. 
This can be attributed to the structure of these games, where most control statements are simple and inherently lead to high branch coverage when optimizing for statement coverage. Additionally, certain branches behave like sequential statements, making explicit branch optimization less impactful. Since $N_{BC}$ systematically explores both true and false paths, it may require additional iterations to fully optimize coverage, whereas $N_{SC}$ achieves similar results more efficiently.


\paragraph*{Insights from Branch Coverage Analysis\\ }
The empirical evaluation of Neatest with Statement Coverage ($N_{SC}$) and Neatest with Branch Coverage ($N_{BC}$) demonstrates that optimizing for branch coverage generally improves test coverage, but with some variability depending on game structure. $N_{BC}$ achieves a higher average coverage of 86.31\% compared to 84.30\% for $N_{SC}$, indicating an improvement in branch coverage when explicitly optimizing for it. However, the extent of this improvement is influenced by factors such as mutually exclusive if-conditions, game randomness, and the complexity of control structures.

For games with intricate decision logic, $N_{BC}$ provides a clear advantage by systematically exploring both true and false branches, ensuring higher fault exposure. However, in games where branches behave similarly to statements, $N_{BC}$’s advantage diminishes. Additionally, highly randomized games can introduce inconsistent performance, resulting in cases where $N_{SC}$ occasionally achieves comparable or even slightly better coverage.

Overall, $N_{BC}$ proves to be a more effective test generation strategy when branch coverage is a critical testing criterion, but its effectiveness depends on the structural characteristics of the software under test.

\subsubsection{Statement Coverge}

\begin{figure}[t]
    \centering
    \includegraphics[width=0.9\linewidth]{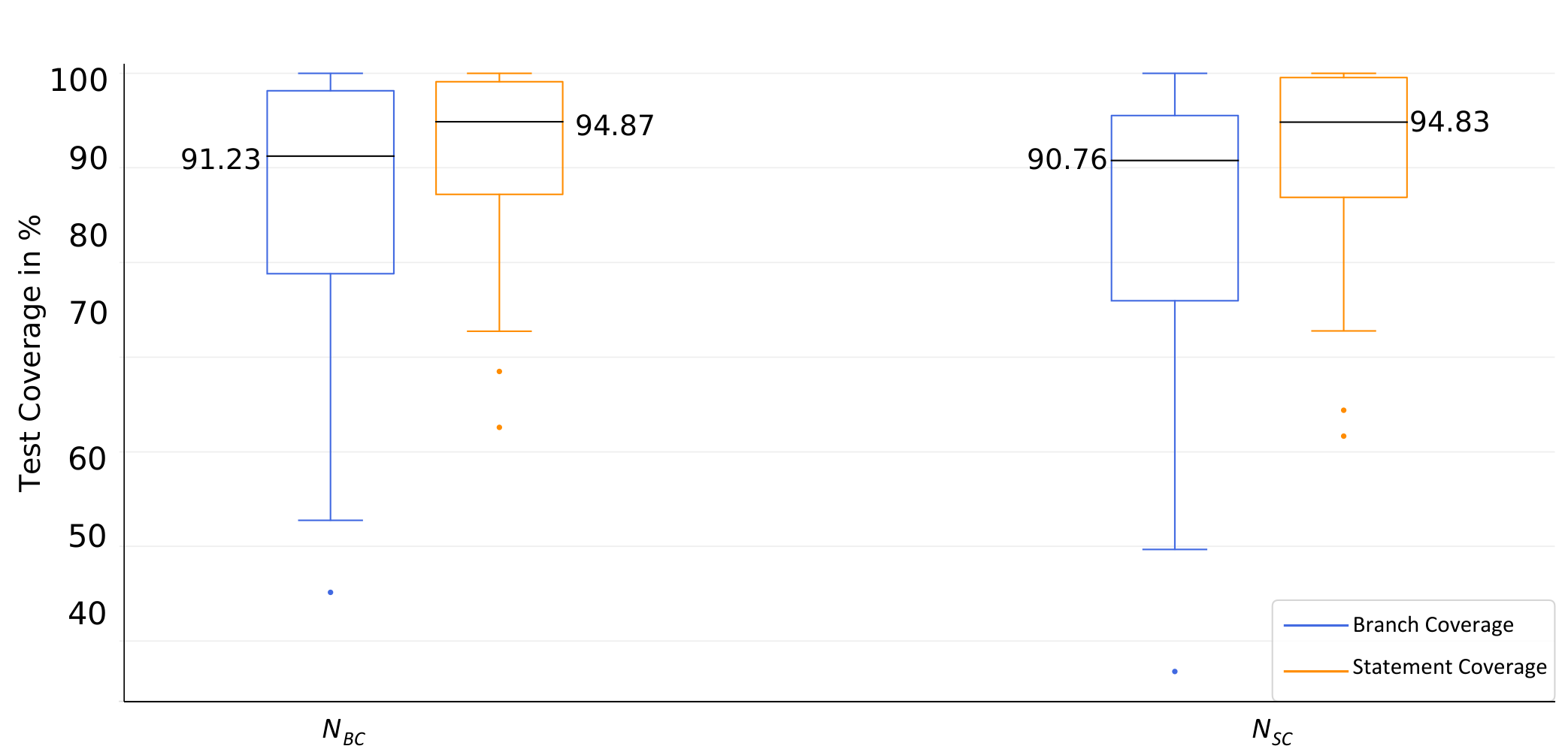}
\caption{Average Branch and Statement coverage percentages for 30 runs of all case studies combined with respect to $N_{BC}$ and $N_{SC}$}
    \label{fig:cov}
\end{figure}

The average Statement coverage percentages obtained for $N_{SC}$ and $N_{BC}$ across 30 experimental runs for each game are summarized in Table~\ref{tab:statement_coverage} and shown in figure \ref{fig:cov}. 
The results indicate that $N_{SC}$ achieved a mean statement coverage of 90.67\%, whereas $N_{BC}$ achieved 91.11\%, demonstrating a slight improvement. However, deeper examination reveals interesting trends based on the structural characteristics of individual games and performance of both $N_{SC}$ and $N_{BC}$ in terms of Statistical significance.

\paragraph*{Case 1: $N_{BC}$ Significantly Outperforms $N_{SC}$ \\}

$N_{BC}$ significantly outperforms $N_{SC}$   in five games (as shown in Table~\ref{tab:statement_coverage}), as indicated by statistically significant p-values ($p < 0.05$). These games include:

The structural complexity of these games explains $N_{BC}$’s superior performance. EndlessRunner and WhackAMole contain deeply nested control structures (e.g., if statements nested up to 4–5 levels). Since control structures introduce multiple statements, achieving high statement coverage requires triggering them at least once.
WhackAMole, SpaceOdyssey, EndlessRunner, OceanCleanup, and DieZauberlehrlinge.

\begin{table}[t]
    \centering

    \footnotesize
    \setlength{\tabcolsep}{3pt} 
    \renewcommand{\arraystretch}{1.1} 
    \caption{Mean Statement coverage\% for \(N_{SC}\) and \(N_{BC}\) for each case study for 30 independent runs along with p-values showing significance and effect size}
    \label{tab:statement_coverage}
    \scriptsize
    \begin{tabular}{|c|l|c|c|c|c|}
       \hline
        \rowcolor{gray!30} \textbf{\#} & \textbf{Project} & \multicolumn{4}{c|}{\textbf{Statement Coverage}} \\
        \cline{3-6}
        \rowcolor{gray!30} & & \textbf{$N_{SC}$} & \textbf{$N_{BC}$} & \textbf{$p$ value} & \textbf{$A12$} \\
        \hline
        \rowcolor{orange!70} 1 & WhackAMole & 75.65 & 78.65 & 0.01 & 0.32 \\
        \rowcolor{orange!70} 2 & SpaceOdyssey & 86.90 & 87.21 & 0.01 & 0.32 \\
        \rowcolor{orange!70} 3 & EndlessRunner & 74.91 & 80.83 & 0.01 & 0.37 \\
        \rowcolor{orange!70} 4 & OceanCleanup & 82.93 & 85.03 & 0.03 & 0.36 \\
        \rowcolor{orange!70} 5 & DieZauberlehrlinge & 64.83 & 68.20 & 0.04 & 0.35 \\
        \rowcolor{green!30} 6 & Dodgeball & 94.83 & 93.68 & 0.03 & 0.57 \\
        \rowcolor{green!30} 7 & Frogger & 93.08 & 93.65 & 0.04 & 0.56 \\
        \rowcolor{green!30} 8 & DesertRally & 93.90 & 93.65 & 0.04 & 0.56 \\
        \rowcolor{green!30} 9 & CityDefender & 72.78 & 75.00 & 0.03 & 0.37 \\
        \rowcolor{gray!70} 10 & BirdsShooter & 100 & 100 & 1.0 & 0.5 \\
        \rowcolor{gray!70} 11 & CatchTheGifts & 100 & 100 & 1.0 & 0.5 \\
        \rowcolor{gray!70} 12 & LineUp & 100 & 100 & 1.0 & 0.5 \\
        \rowcolor{gray!70} 13 & BrainGame & 100 & 100 & 1.0 & 0.5 \\
        \rowcolor{gray!70} 14 & SnowballFight & 95.21 & 95.21 & 0.5 & 0.5 \\
        \rowcolor{gray!70} 15 & CatchingApples & 100 & 100 & 1.0 & 0.5 \\
        \rowcolor{gray!70} 16 & RioShootout & 100 & 100 & 1.0 & 0.5 \\
        \rowcolor{blue!30} 17 & FlappyParrot & 91.35 & 92.07 & 0.16 & 0.46 \\
        \rowcolor{blue!30} 18 & Snake & 91.72 & 94.80 & 0.15 & 0.47 \\
        \rowcolor{blue!30} 19 & HackAttack & 92.97 & 93.65 & 0.25 & 0.48 \\
        \rowcolor{gray!90} 20 & Dragons & 100 & 100 & 1.0 & 0.5 \\
        \rowcolor{gray!90} 21 & CatchTheDots & 99.55 & 98.91 & 0.25 & 0.39 \\
        \rowcolor{gray!90} 22 & FallingStars & 98.87 & 97.65 & 0.3 & 0.45 \\
        \rowcolor{gray!90} 23 & FinalFight & 98.97 & 98.07 & 0.5 & 0.53 \\
        \rowcolor{gray!90} 24 & Pong & 98.97 & 98.07 & 0.5 & 0.53 \\
        \rowcolor{gray!90} 25 & Fruitcatching & 96.67 & 96.18 & 0.75 & 0.52 \\
        \hline
        \rowcolor{gray!40} & \textbf{Mean} & 90.67 & 91.11 & - & 0.4675 \\
        \hline
    \end{tabular}
\end{table}
$N_{SC}$ explicitly targets uncovered statements but does not prioritize executing alternative paths within a branching structure. As a result, statements within unexecuted branches may remain uncovered. Conversely, $N_{BC}$ systematically ensures that both paths of a conditional statement (if and else) are exercised, increasing the likelihood of reaching additional statements.

This advantage is evident in OceanCleanup and DieZauberlehrlinge, which contain deeply nested if-else structures (up to six levels), and SpaceOdyssey, where one sprite has three levels of nested conditions. Since $N_{SC}$ does not enforce execution of both sides of a condition, some statements remain uncovered. $N_{BC}$, by ensuring structured exploration, naturally improves statement coverage.

While $N_{BC}$ consistently achieves higher mean coverage, statistical validation confirms that this improvement is systematic rather than incidental. The Mann-Whitney U test determines whether one approach consistently outperforms the other across multiple runs.

\begin{itemize}
    \item A low p-value ($p < 0.05$) confirms that $N_{BC}$’s higher statement coverage is statistically significant.
    \item Higher mean coverage alone does not imply significance; consistency across experimental runs determines superiority.
\end{itemize}

For these five games, $N_{BC}$’s improved statement coverage is reinforced by low p-values, confirming that its structured approach leads to consistent gains.

\paragraph*{Case 2: $N_{BC}$ Achieves Higher Mean Coverage, without Statistical Significance \\ }
In four games, $N_{BC}$ achieved better mean coverage than $N_{SC}$ (as shown in Table~\ref{tab:statement_coverage}), but the p-values exceeded 0.05, indicating high variance in performance across runs. These games are:
FlappyParrot, Snake, HackAttack, and Dragons.

The inconsistency in $N_{BC}$’s performance can be attributed to variability in evolved test strategies. Additionally, static analysis of the game structures reveals that several if-conditions are mutually exclusive, meaning that achieving one condition automatically negates the other.
For example, in FlappyParrot and Snake, the primary control structures are forever loops and halting blocks. These blocks are important for achieving statement coverage. Hence, NSC has more reliable performance throughout various runs.
In HackAttack, the game structure includes score-dependent and chance-based termination conditions, leading to inconsistent NBC performance. The Virus sprite interacts with Neo-cat to increase the score, but the game halts at a fixed threshold. Additionally, the loss condition depends on a separate counter reducing over time. These constraints limit NBC’s exploration, making its performance highly variable.
In the Dragons game, NBC’s search process requires more sophisticated evolution strategies to optimize coverage.

\paragraph*{Case 3: $N_{BC}$ and $N_{SC}$ Achieve Equal Coverage \\ }
In 7 games, $N_{SC}$ and $N_{BC}$ achieve nearly identical statement coverage (as shown in Table~\ref{tab:statement_coverage}). These games are:
 BirdShooter, CatchTheGifts, LineUp, BrainGame, SnowballFight, CatchingApples, and RioShootout. 

This outcome is primarily influenced by the simplicity of the game logic and the presence of structural constraints that limit potential differences between the two approaches. In BirdShooter and CatchingApples, the code structure is straightforward, allowing both approaches to easily achieve full coverage. Prior results from RQ1.1 (Table \ref{tab:branch_coverage}) demonstrated that even branch coverage was maximized in these games, reinforcing the notion that their logic does not impose significant challenges for either technique.
For SnowballFight and RioShootout, dead code prevents either method from reaching full statement coverage. Since these unreachable statements remain unexecuted regardless of the testing strategy, both approaches exhibit similar performance.
In CatchTheGifts, LineUp, and BrainGame, while control structures exist, executing just one control path is sufficient to achieve high statement coverage. Unlike games with deeply nested branching logic, these structures do not demand systematic exploration of both paths to maximize coverage. As a result, NSC performs comparably to NBC, despite its probabilistic approach to statement execution.

\paragraph*{Case 4: $N_{SC}$ Achieves Higher Mean Coverage, without Statistical Significance \\ }
In 5 games, $N_{SC}$ achieved slightly higher mean coverage than $N_{BC}$ (as shown in Table~\ref{tab:statement_coverage}). These games are:
CatchTheDots, FallingStars, FinalFight, Pong, and FruitCatching.

Although the differences were minimal (ranging from 0\% to 1.18\%), this trend suggests that $N_{SC}$ performs better in games with fewer and simpler control dependencies or a higher degree of randomness.
In CatchTheDots and FruitCatching, the number of statements is low, and the control structures are simple. Many conditional statements in these games, such as if or else-if blocks, are mutually exclusive. This means that triggering one condition (e.g., a true path) inherently implies coverage of the contrasting condition (e.g., a false path). For example, in FruitCatching, if a fruit falls into the bowl (true path), it does not reach the ground (false path), allowing both statements to be covered through a single execution path. Similarly, in CatchTheDots, if a red ball appears and is caught by the red controller, the score increases by one; otherwise, a life is deducted. Since these conditions are mutually exclusive, NSC efficiently covers them.
Pong has the lowest number of statements and sprites, making full coverage straightforward for NSC.
FinalFight contains a high number of statements but very few control structures, minimizing the advantage of NBC’s systematic branching approach.
Lastly, in FallingStars, intrinsic randomness can cause NSC to incidentally trigger more if conditions than NBC, contributing to its slightly better performance in statement coverage.

\paragraph*{Case 5: $N_{SC}$ Significantly Outperforms $N_{BC}$ \\ }
In four games, Dodgeball, Frogger, DesertRally, and CityDefender, $N_{SC}$ attains slightly higher mean statement coverage (as shown in Table~\ref{tab:statement_coverage}). 
The statistical significance of the p-values \((p < 0.05)\) confirms that $N_{SC}$'s superior performance in this case is not due to random variation but a consistent trend across multiple runs. 

This can be attributed to the inherent focus of $N_{SC}$ on maximizing executed statements, leading to more efficient coverage in games where most control structures do not significantly alter execution paths. Additionally, the presence of mutually exclusive conditions and simple branching structures may allow $N_{SC}$  to achieve near-complete coverage without the need for extensive exploration. Since 
$N_{BC}$ prioritizes branch execution, it may allocate resources inefficiently in cases where achieving full statement coverage does not necessarily require optimizing for both true and false branches.

\paragraph*{Insights from from Statement Coverage \\ }

The comparative analysis of 
$N_{SC}$ and $N_{BC}$ in statement coverage highlights several key insights. In structurally complex games with deeply nested control structures, such as WhackAMole, SpaceOdyssey, and EndlessRunner, $N_{BC}$ consistently achieves higher statement coverage due to its ability to systematically explore all conditional branches. Conversely, in games with simpler control flows, such as Dodgeball, Frogger, and CityDefender, $N_{SC}$  performs better by maximizing executed statements without the need for exhaustive branch exploration. Notably, in cases where both approaches achieve similar coverage, such as BirdShooter, LineUp, and BrainGame, the game logic is either straightforward or contains dead code, limiting the potential differences between the two techniques. Additionally, in certain games like FlappyParrot and Snake, the presence of mutually exclusive conditions results in high variance for $N_{BC}$, leading to inconsistent performance. Overall, while $N_{BC}$ demonstrates an advantage in structurally intricate games, $N_{SC}$ remains more efficient in scenarios where statement coverage can be achieved without enforcing both conditional paths.

\begin{tcolorbox}[colback=gray!20,colframe=gray!60,title=\textbf{RQ1: Test Coverage Comparison: $N_{BC}$ vs. $N_{SC}$}]


$N_{BC}$ provides better branch and statement coverage in structurally intricate games, while $N_{SC}$ is more efficient for straightforward control flows. 

\end{tcolorbox}

\subsection{RQ2: Mutation Analysis for Test Oracle Effectiveness}
\label{ssec:rq2}
To evaluate the reliability of $N_{BC}$ and $N_{SC}$ as test oracles, we conducted a mutation analysis on 205,545 mutants across 25 games using eight different mutation operators. The objective was to assess how well each method differentiates between true faults and randomized program behavior, considering both mutation kill scores and false positive rates.

\begin{table*}[t]
    \centering
    
    \footnotesize
    \caption{Mutation kill score \% achieved by $N_{SC}$ and $N_{BC}$ }
    \label{tab:mutation_score}
    \scriptsize
    \renewcommand{\arraystretch}{1}
    \begin{tabular}{|l|p{1.5cm}|p{1.5cm}|c|p{1.5cm}|p{1.5cm}|c|}
        \hline
        \rowcolor{white} 
        \textbf{Projects} & \textbf{\#Total mutants generated for $N_{SC}$} & \textbf{Average Mutation Score $N_{SC}$} & \textbf{\#FP} & \textbf{\#Total mutants generated for $N_{BC}$}  & \textbf{Average Mutation Score $N_{BC}$} & \textbf{\#FP} \\
        \hline
        \rowcolor{orange!30} FlappyParrot & 1410 & 58.66 & 1.0 & 1410 & 61.66 & 1.0 \\
        \rowcolor{orange!30} Frogger & 6110 & 95.63 & 3.00 & 6045 & 95.86 & 2.78 \\
        \rowcolor{orange!30} Fruitcatching & 2549 & 90.25 & 13.44 & 2544 & 91.58 & 15.00 \\
        \rowcolor{orange!30} CatchTheDots & 3734 & 84.65 & 7.22 & 3745 & 91.80 & 14.29 \\
        \rowcolor{orange!30} CatchTheGifts & 3334 & 85.47 & 69.51 & 3331 & 87.14 & 59.74 \\
        \rowcolor{orange!30} EndlessRunner & 6690 & 57.05 & 2.66 & 6690 & 60.94 & 12.00 \\
        \rowcolor{orange!30} CatchingApples & 1290 & 79.89 & 35.33 & 1290 & 83.96 & 35.66 \\
        \rowcolor{orange!30} FallingStars & 5520 & 92.50 & 29.27 & 5529 & 93.31 & 26.51 \\
        \rowcolor{orange!30} SpaceOdyssey & 3780 & 64.16 & 26.33 & 3792 & 66.70 & 29.66 \\
        \rowcolor{orange!30} SnowballFight & 1830 & 63.83 & 3.33 & 1830 & 77.92 & 3.00 \\
        \rowcolor{orange!30} DieZauberlehrlinge & 4959 & 38.89 & 1.00 & 4955 & 43.17 & 3.66 \\
        \rowcolor{orange!30} Dodgeball & 3108 & 95.12 & 29.48 & 3102 & 96.07 & 23.40 \\
        \hline
        \rowcolor{green!30} FinalFight & 8533 & 69.14 & 45.68 & 8498 & 66.40 & 36.25 \\
        \rowcolor{green!30} WhackAMole & 8957 & 90.83 & 72.07 & 9006 & 87.50 & 58.40 \\
        \rowcolor{green!30} BrainGame & 2632 & 51.76 & 20.00 & 2633 & 49.18 & 15.33 \\
        \rowcolor{green!30} DesertRally & 6960 & 100 & 11.66 & 6974 & 99.92 & 9.33 \\
        \hline
        \rowcolor{gray!30} OceanCleanup & 4800 & 91.30 & 62.33 & 4800 & 91.03 & 62.00 \\
        \rowcolor{gray!30} Pong & 390 & 68.22 & 27.66 & 390 & 64.44 & 28.92 \\
        \rowcolor{gray!30} RioShootout & 4701 & 43.48 & 4.00 & 4710 & 41.17 & 3.00 \\
        \rowcolor{gray!30} Dragons & 7200 & 40.17 & 5.16 & 7295 & 33.91 & 5.35 \\
        \rowcolor{gray!30} HackAttack & 2842 & 83.68 & 17.37 & 2847 & 80.59 & 15.07 \\
        \rowcolor{gray!30} LineUp & 2340 & 72.95 & 42.00 & 2340 & 64.53 & 35.33 \\
        \rowcolor{gray!30} BirdsShooter & 2907 & 92.02 & 14.81 & 2907 & 90.56 & 16.22 \\
        \rowcolor{gray!30} Snake & 3015 & 59.60 & 10.0 & 3017 & 48.93 & 2.66 \\
        \rowcolor{gray!30} CityDefender & 3112 & 45.09 & 57.7 & 3162 & 45.72 & 1.69 \\
        \hline
        \rowcolor{gray!15} Mean & 4108.12 & 72.57 & 22.40 & 4113.68 & 72.15 & 20.65 \\
        \hline
    \end{tabular}
    
\end{table*}

\begin{figure*}[t]
    \centering
    \includegraphics[width=\linewidth]{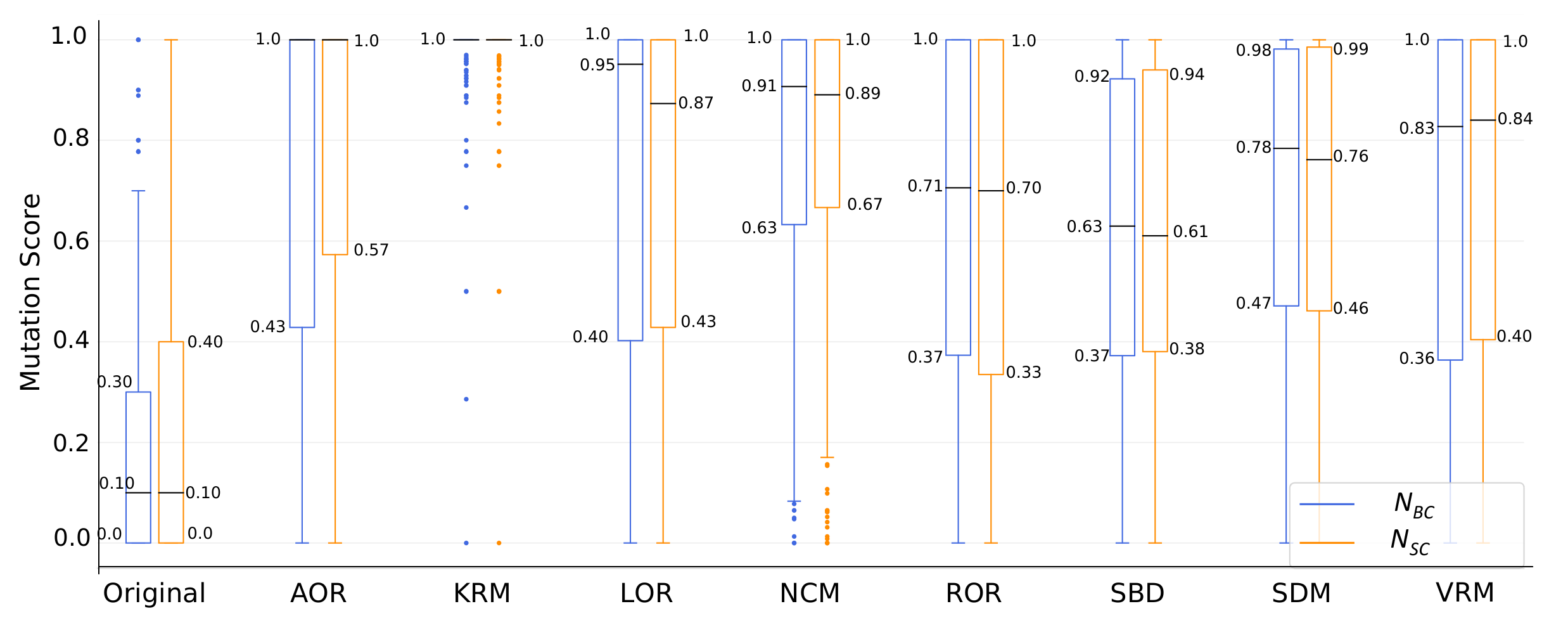}
\caption{Average mutation scores for 30 runs of all case studies combined with respect to each mutation operator for $N_{BC}$ and $N_{SC}$}
    \label{fig:mutation}
\end{figure*}

The results, summarized in Table~\ref{tab:mutation_score}, provide insight into the precision of each approach. Figure \ref{fig:mutation} shows average mutation scores for 30 runs of all case studies combined with respect to each mutation operator for $N_{BC}$ and $N_{SC}$.

The average mutation score for $N_{SC}$ was 72.57\%, slightly higher than $N_{BC}$’s 72.15\%. However, $N_{SC}$ also exhibited a higher false positive rate of 22.40\%, whereas $N_{BC}$ maintained a lower false positive rate of 20.65\%. While the difference in mean mutation scores is small (0.42\%), the difference in false positives (1.75\%) suggests that $N_{SC}$ tends to over-detect faults. This implies that $N_{SC}$ often misclassifies normal program variations as faults, making it less precise. Conversely, $N_{BC}$, with its lower false positive rate, demonstrates a more refined ability to detect actual faults while reducing unnecessary false alarms.

\paragraph*{Case 1: $N_{BC}$ Detects More Faults with Fewer False Positives \\}
$N_{BC}$ showed superior performance in several games where decision-making logic played a significant role in the program's execution (as shown in Table~\ref{tab:mutation_score}). These include FlappyParrot, Frogger, FruitCatching, EndlessRunner, CatchTheDots, CatchTheGifts, CatchingApples, FallingStars, SpaceOdyssey, SnowballFight, DieZauberlehrlinge, and Dodgeball.

In these cases, $N_{BC}$ successfully identified a greater number of true faults while maintaining a lower rate of misclassification. The results indicate that optimizing for branch coverage allows $N_{BC}$ to evolve networks that are more effective at distinguishing between faulty and non-faulty behavior. This suggests that test cases generated with a branch coverage objective are better suited for exposing control-dependent errors, particularly in decision-heavy programs.

\paragraph*{Case 2: $N_{SC}$ Appears to Outperform $N_{BC}$, due to False Positives \\}
In contrast, there were also a few cases where $N_{SC}$ appeared to outperform $N_{BC}$ in mutation detection (as shown in Table~\ref{tab:mutation_score}), particularly in LineUp, BrainGame, Snake, CityDefender, DesertRally, FinalFight, OceanCleanup, Pong, RioShootout, Dragons, WhackAMole, HackAttack, and BirdsShooter.

However, closer examination of the data reveals that $N_{SC}$’s higher mutation kill scores were largely driven by its tendency to generate more false positives, rather than detecting a greater number of real faults. This means that while $N_{SC}$ may appear to be detecting more faults, it is in fact failing to differentiate between expected variations in randomized program execution and actual software defects.

For instance, in Pong and RioShootout, $N_{SC}$ achieved a higher kill score, but this was due to misclassifying random execution variations as genuine faults, inflating its overall score. In FinalFight, $N_{SC}$ detected a slightly greater number of faults, but also misclassified a significantly larger number of non-faulty behaviors, further demonstrating its tendency to over-predict errors.

\paragraph*{Case 3: Performance Across Mutation Operators\\}
Further analysis of mutation operators shows that $N_{BC}$ consistently produced lower standard deviations (10\%) compared to $N_{SC}$ (30\%), indicating a more stable and reliable fault detection mechanism (as depicted in Figure \ref{fig:mutation}). In Arithmetic Operator Replacement (AOR), both methods achieved a 100\% mutation kill rate, but $N_{BC}$’s higher node density suggests a more stable execution process.

$N_{BC}$ also demonstrated greater effectiveness in detecting faults in decision structures, particularly in Negate Conditional Mutation (NCM) and Relational Operator Replacement (ROR), where it outperformed $N_{SC}$ with higher standard deviations of 91\% and 71\%, respectively. In Single Block Deletion (SBD) and Script Deletion Mutation (SDM), $N_{BC}$ similarly achieved superior results.

However, Variable Replacement Mutation (VRM) posed a challenge for $N_{BC}$, as its mutation score was 83\% with higher density, indicating that variable mutations do not always impact control structures unless they alter decision-making processes. This suggests that $N_{SC}$, which focuses on executing every statement at least once, may have an advantage in detecting faults in non-decision variables, even though it lacks $N_{BC}$’s precision in identifying control-flow-related faults.

\paragraph*{Insights from from Mutation Analysis\\}
The results of this analysis highlight that while $N_{SC}$ may achieve slightly higher overall mutation kill scores, this advantage is offset by its higher false positive rate, making it less reliable as a test oracle. $N_{BC}$, despite marginally lower kill scores, proves to be a more precise method for mutation testing, as it effectively reduces the likelihood of misidentifying non-faulty variations as defects.

The inherent randomness of Scratch games further amplifies $N_{SC}$’s shortcomings, as it frequently mislabels randomized behaviors as faults, leading to misleading conclusions. While $N_{SC}$ may still be useful in detecting a broad range of faults, its higher false positive rate reduces confidence in its results, making $N_{BC}$ the preferred choice for scenarios where accuracy and fault differentiation are critical.

Ultimately, $N_{BC}$ emerges as the more robust and precise approach for fault detection in Scratch programs. Its ability to prioritize branch exploration and distinguish real faults from expected program behavior makes it a better-suited testing strategy for complex, decision-based Scratch applications.

\begin{tcolorbox}[colback=gray!20,colframe=gray!60,title=\textbf{RQ2: Test Oracle Effectiveness}]
$N_{BC}$ is a more robust and precise test oracle, especially for complex, decision-based Scratch applications, as it effectively reduces false positives while maintaining strong fault detection capabilities.
\end{tcolorbox}
\subsection{Discussion and Analysis}

The empirical results presented in subsections~\ref{ssec:rq1} and ~\ref{ssec:rq2} provide key insights into the performance differences between Neatest with Statement Coverage ($N_{SC}$) and Neatest with Branch Coverage ($N_{BC}$). This subsection discusses the implications of these findings, analyzes the strengths and weaknesses of each approach, and outlines directions for future research.

\subsubsection{Coverage Performance and Structural Impact}

Our findings demonstrate that $N_{BC}$ achieves higher mean branch coverage (86.31\%) compared to $N_{SC}$ (84.30\%), but the degree of improvement varies significantly depending on the structural complexity of the Scratch program.
Games with rich control dependencies (e.g., conditional blocks with multiple decision paths) benefited the most from $N_{BC}$. This is because $N_{BC}$ actively prioritizes branch execution, ensuring that both the true and false paths of conditions are exercised, leading to better fault detection.
Conversely, games with mutually exclusive branches (i.e., control structures that inherently cancel each other out) showed no substantial improvement with $N_{BC}$ over $N_{SC}$. In such cases, statement execution was sufficient to achieve high coverage without the need for explicit branch targeting.
This suggests that the effectiveness of $N_{BC}$ is highly dependent on program structure, meaning that blindly applying branch coverage as a default metric may not always lead to optimal test generation results.

\subsubsection{Test Oracle Effectiveness: Mutation Kill Score vs. False Positives}

When assessing the ability of test cases to act as oracles, we found that while $N_{SC}$ achieved a slightly higher mutation kill score (72.57\%) compared to $N_{BC}$ (72.15\%), it also exhibited a significantly higher false positive rate (22.40\%) compared to $N_{BC}$’s 20.65\%.

$N_{BC}$’s lower false positive rate suggests that it is better at distinguishing between true faults and random program behavior.
$N_{SC}$, while achieving higher overall kill scores, often misclassified correct behaviors as faults, leading to an inflation of its mutation detection results.
This highlights an important trade-off: $N_{BC}$ provides a more precise fault detection mechanism, whereas $N_{SC}$ may produce misleadingly high kill scores due to over-sensitive fault detection.

\subsubsection{Impact of Randomness in Scratch Games}

One key observation from our experiments is that randomness in game behavior significantly influences test generation effectiveness.

For games with high randomness (e.g., FallingStars, CityDefender, and Snake), neither $N_{SC}$ nor $N_{BC}$ was able to consistently outperform the other.
In such games, the inherent variability in program execution makes it difficult to optimize test cases, since the same sequence of inputs may lead to different program states.
This raises an important consideration: test generation strategies for highly stochastic environments may require additional mechanisms, such as reinforcement learning or adaptive heuristics, to improve reliability.

\subsubsection{Practical Implications for Automated Game Testing}

From a practical standpoint, our results suggest that branch coverage should not always be treated as the default metric for game testing.

For structurally simple games, $N_{SC}$ is often sufficient to achieve high test coverage with lower computational overhead.
For complex games involving decision trees, event-driven behaviors, or nested control structures, $N_{BC}$ provides a meaningful advantage by ensuring that all possible execution paths are explored.
Test oracle effectiveness is critical—$N_{BC}$'s lower false positive rate makes it more suitable for applications where distinguishing actual faults from random behavior is essential.
Future work should explore hybrid approaches, where test generation dynamically switches between statement and branch coverage strategies based on program complexity. Additionally, context-aware fitness functions could be integrated to adaptively adjust search criteria based on game state variability.




Our results suggest that $N_{BC}$ generally outperforms $N_{SC}$ for complex Scratch programs, particularly when full logical path coverage is required. Additionally, $N_{BC}$ provides more precise fault detection with fewer false positives. Future research should focus on developing hybrid and adaptive test generation approaches to further improve automated game testing for Scratch programs.

\section{Threats to Validity}
\label{sec:threats}

This section outlines the potential threats to validity in our study, categorized into external validity, internal validity, and construct validity.

\subsection{External Validity}
External validity concerns the generalizability of our findings beyond the specific dataset used in this study. Our dataset includes 25 diverse Scratch games, carefully selected to represent a wide range of game structures, interaction mechanisms, and control dependencies. However, we cannot guarantee that our results generalize to all Scratch applications, particularly large-scale commercial projects or games developed in different programming languages. Expanding the dataset in future studies to include a broader variety of Scratch projects and non-game applications could enhance the generalizability of our findings.

\subsection{Internal Validity}
Internal validity relates to factors that could affect the accuracy and reliability of our experimental results. One key limitation is the randomized nature of the experiments, which may cause slight variations in results across different runs. We mitigated this threat by performing 30 repetitions for RQ1 (test coverage evaluation) and 100 repetitions for RQ2 (mutation analysis) to ensure statistical robustness.

Additionally, our experiments used a fixed search budget of 5 hours per test case. While this was sufficient for most Scratch games, highly complex games with extensive control dependencies may require longer execution times for full test suite optimization. Dynamic search budgets, based on game complexity, could be explored in future work to improve coverage for more complex Scratch applications.

Furthermore, the inherent randomness in Scratch game execution posed challenges, particularly for games with stochastic behaviors. Although we accounted for this by averaging results across multiple runs, adaptive testing strategies that dynamically adjust search techniques based on program state variability could further enhance test generation effectiveness.

\subsection{Construct Validity}
Construct validity pertains to whether our study accurately measures the intended concepts. In our evaluation, we used branch coverage, statement coverage, and mutation analysis as primary metrics. While these metrics are widely accepted in software testing research, they may not fully capture all dimensions of test effectiveness, such as real-world bug detection capabilities.

Additionally, the false positives observed in mutation analysis could slightly skew the perceived effectiveness of $N_{SC}$ and $N_{BC}$ as test oracles. $N_{SC}$’s higher mutation kill scores were sometimes due to over-sensitive fault detection, leading to more false positives. Future work could incorporate additional evaluation criteria, such as developer feedback on test case utility or real-world bug detection studies.

\section{Conclusion}
\label{sec:conclusion}
Automated test generation for game-like programs presents unique challenges due to their intentional complexity, dynamic user interactions, and inherent randomness. Traditional static test generators struggle to effectively test these programs as they fail to differentiate between novel and erroneous behaviors. This limitation prevents them from achieving comprehensive test coverage and robust fault detection. To address these challenges, this study introduced a novel test generator with a branch coverage-focused fitness function, extending the capabilities of Neatest for Scratch game testing.

Our approach leverages neuroevolution-based dynamic test suites, where neural networks evolve to reliably reach program states, including winning conditions, despite randomized execution paths. By incorporating branch coverage optimization, our method ensures higher test coverage, overcoming the limitations of previous test generators that only optimized for statement coverage. Empirical evaluation across 25 diverse Scratch games demonstrates that our branch-coverage-driven approach achieves higher test coverage ratios and improved fault detection capabilities compared to statement coverage-based test generation. Additionally, the networks function as test oracles, leveraging their structure to detect unexpected behaviors in mutant programs. The results show that Neatest with Branch Coverage ($N_{BC}$) consistently achieves more reliable test coverage while reducing false positives, making it a more effective approach for automated game testing.

While our approach significantly improves coverage and fault detection, several areas remain open for future research. Expanding the dataset beyond the 25 Scratch games used in this study, including commercial Scratch projects and non-game educational applications, would provide broader validation of our method. Another key direction is adaptive search strategies, as our fixed search budget of five hours per test case may not be optimal for highly complex games. Future research should explore dynamic resource allocation based on game complexity and execution behavior. Additionally, handling non-deterministic behaviors remains a challenge, as games with highly randomized execution paths require more advanced test generation techniques. Integrating reinforcement learning-based strategies or adaptive heuristics could further enhance test generation stability in stochastic environments.

Furthermore, our current approach focuses solely on branch coverage, but future work should explore hybrid test generation techniques that dynamically switch between branch and statement coverage based on program structure and testing needs. This could further improve test efficiency and fault detection rates. Another promising research avenue is real-world bug detection and developer feedback integration, where the effectiveness of $N_{BC}$-generated test cases is assessed based on actual defect discovery in Scratch programs. Finally, integrating our method into educational and industry-level game testing frameworks could provide practical validation and real-world applicability, especially for large-scale game development environments like Unity or Godot.

In summary, this research advances automated game testing by enhancing coverage criteria, fault detection, and test oracle effectiveness. By leveraging neuroevolution and dynamic test suites, our approach bridges the gap between automated test generation and human-like gameplay interactions. While $N_{BC}$ clearly outperforms statement coverage-based methods, further research on adaptive search strategies, stochastic behavior handling, and real-world applicability will further solidify its role as a robust automated testing framework for game-like programs.



%
%

\bibliography{Sources.bib}   

\begin{thebibliography}{10}

\bibitem{statista2025games}
{Statista}.
\newblock Games - worldwide.
\newblock \url{https://www.statista.com/outlook/amo/media/games/worldwide?currency=USD}, 2025.
\newblock Accessed: 2025-05-02.

\bibitem{deiner2023automated}
Adina Deiner, Patric Feldmeier, Gordon Fraser, Sebastian Schweikl, and Wengran Wang.
\newblock Automated test generation for scratch programs.
\newblock {\em Empirical Software Engineering}, 28(3):79, 2023.

\bibitem{politowski2021survey}
Cristiano Politowski, Fabio Petrillo, and Yann-Ga{\"e}l Gu{\'e}h{\'e}neuc.
\newblock A survey of video game testing.
\newblock In {\em 2021 IEEE/ACM International Conference on Automation of Software Test (AST)}, pages 90--99. IEEE, 2021.

\bibitem{diah2010usability}
Norizan~Mat Diah, Marina Ismail, Suzana Ahmad, and Mohd Khairulnizam~Md Dahari.
\newblock Usability testing for educational computer game using observation method.
\newblock In {\em 2010 international conference on information retrieval \& knowledge management (CAMP)}, pages 157--161. IEEE, 2010.

\bibitem{ferre2009playability}
Xavier Ferre, Angelica de~Antonio, Ricardo Imbert, and Nelson Medinilla.
\newblock Playability testing of web-based sport games with older children and teenagers.
\newblock In {\em Human-Computer Interaction. Interacting in Various Application Domains: 13th International Conference, HCI International 2009, San Diego, CA, USA, July 19-24, 2009, Proceedings, Part IV 13}, pages 315--324. Springer, 2009.

\bibitem{cho2010online}
Chang-Sik Cho, Kang-Min Sohn, Chang-Jun Park, and Ji-Hoon Kang.
\newblock Online game testing using scenario-based control of massive virtual users.
\newblock In {\em 2010 The 12th International Conference on Advanced Communication Technology (ICACT)}, volume~2, pages 1676--1680. IEEE, 2010.

\bibitem{schaefer2013crushinator}
Christopher Schaefer, Hyunsook Do, and Brian~M Slator.
\newblock Crushinator: A framework towards game-independent testing.
\newblock In {\em 2013 28th IEEE/ACM International Conference on Automated Software Engineering (ASE)}, pages 726--729. IEEE, 2013.

\bibitem{smith2009computational}
Adam Smith, Mark Nelson, and Michael Mateas.
\newblock Computational support for play testing game sketches.
\newblock In {\em Proceedings of the AAAI Conference on Artificial Intelligence and Interactive Digital Entertainment}, volume~5, pages 167--172, 2009.

\bibitem{ariyurek2019automated}
Sinan Ariyurek, Aysu Betin-Can, and Elif Surer.
\newblock Automated video game testing using synthetic and humanlike agents.
\newblock {\em IEEE Transactions on Games}, 13(1):50--67, 2019.

\bibitem{pfau2017automated}
Johannes Pfau, Jan~David Smeddinck, and Rainer Malaka.
\newblock Automated game testing with icarus: Intelligent completion of adventure riddles via unsupervised solving.
\newblock In {\em Extended abstracts publication of the annual symposium on computer-human interaction in play}, pages 153--164, 2017.

\bibitem{politowski2022towards}
Cristiano Politowski, Yann-Ga{\"e}l Gu{\'e}h{\'e}neuc, and Fabio Petrillo.
\newblock Towards automated video game testing: still a long way to go.
\newblock In {\em Proceedings of the 6th international ICSE workshop on games and software engineering: engineering fun, inspiration, and motivation}, pages 37--43, 2022.

\bibitem{xue2022learning}
Feng Xue, Junsheng Wu, and Tao Zhang.
\newblock Learning-replay based automated robotic testing for mobile app.
\newblock {\em Mobile Information Systems}, 2022(1):1084602, 2022.

\bibitem{feldmeier2023learning}
Patric Feldmeier and Gordon Fraser.
\newblock Learning by viewing: Generating test inputs for games by integrating human gameplay traces in neuroevolution.
\newblock In {\em Proceedings of the Genetic and Evolutionary Computation Conference}, pages 1508--1517, 2023.

\bibitem{maloney2010scratch}
John Maloney, Mitchel Resnick, Natalie Rusk, Brian Silverman, and Evelyn Eastmond.
\newblock The scratch programming language and environment.
\newblock {\em ACM Transactions on Computing Education (TOCE)}, 10(4):1--15, 2010.

\bibitem{stahlbauer2019testing}
Andreas Stahlbauer, Marvin Kreis, and Gordon Fraser.
\newblock Testing scratch programs automatically.
\newblock In {\em Proceedings of the 2019 27th ACM Joint Meeting on European Software Engineering Conference and Symposium on the Foundations of Software Engineering}, pages 165--175, 2019.

\bibitem{feldmeier2022neuroevolution}
Patric Feldmeier and Gordon Fraser.
\newblock Neuroevolution-based generation of tests and oracles for games.
\newblock In {\em Proceedings of the 37th IEEE/ACM International Conference on Automated Software Engineering}, pages 1--13, 2022.

\bibitem{stanley2002evolving}
Kenneth~O Stanley and Risto Miikkulainen.
\newblock Evolving neural networks through augmenting topologies.
\newblock {\em Evolutionary computation}, 10(2):99--127, 2002.

\bibitem{fraser2012whole}
Gordon Fraser and Andrea Arcuri.
\newblock Whole test suite generation.
\newblock {\em IEEE Transactions on Software Engineering}, 39(2):276--291, 2012.

\bibitem{ntafos1979path}
Simeon~C. Ntafos and S.~Louis Hakimi.
\newblock On path cover problems in digraphs and applications to program testing.
\newblock {\em IEEE Transactions on Software Engineering}, (5):520--529, 1979.

\bibitem{yang1998all}
Cheer-Sun~D Yang, Amie~L Souter, and Lori~L Pollock.
\newblock All-du-path coverage for parallel programs.
\newblock {\em ACM SIGSOFT Software Engineering Notes}, 23(2):153--162, 1998.

\bibitem{graham2008foundations}
Dorothy Graham, Erik~van Veenendaal, Isabel Evans, and Rex Black.
\newblock {\em Foundations of software testing: ISTQB certification}.
\newblock Intl Thomson Business Pr, 2008.

\bibitem{hemmati2015effective}
Hadi Hemmati.
\newblock How effective are code coverage criteria?
\newblock In {\em 2015 IEEE International Conference on Software Quality, Reliability and Security}, pages 151--156. IEEE, 2015.

\bibitem{zhu1997software}
Hong Zhu, Patrick~AV Hall, and John~HR May.
\newblock Software unit test coverage and adequacy.
\newblock {\em Acm computing surveys (csur)}, 29(4):366--427, 1997.

\bibitem{mcminn2004search}
Phil McMinn.
\newblock Search-based software test data generation: a survey.
\newblock {\em Software testing, Verification and reliability}, 14(2):105--156, 2004.

\bibitem{wegener2001evolutionary}
Joachim Wegener, Andr{\'e} Baresel, and Harmen Sthamer.
\newblock Evolutionary test environment for automatic structural testing.
\newblock {\em Information and software technology}, 43(14):841--854, 2001.

\bibitem{korel1990automated}
Bogdan Korel.
\newblock Automated software test data generation.
\newblock {\em IEEE Transactions on software engineering}, 16(8):870--879, 1990.

\bibitem{baresi2010testful}
Luciano Baresi, Pier~Luca Lanzi, and Matteo Miraz.
\newblock Testful: an evolutionary test approach for java.
\newblock In {\em 2010 Third International Conference on Software Testing, Verification and Validation}, pages 185--194. IEEE, 2010.

\bibitem{fraser2011evosuite}
Gordon Fraser and Andrea Arcuri.
\newblock Evosuite: automatic test suite generation for object-oriented software.
\newblock In {\em Proceedings of the 19th ACM SIGSOFT symposium and the 13th European conference on Foundations of software engineering}, pages 416--419, 2011.

\bibitem{tonella2004evolutionary}
Paolo Tonella.
\newblock Evolutionary testing of classes.
\newblock {\em ACM SIGSOFT Software Engineering Notes}, 29(4):119--128, 2004.

\bibitem{arcuri2019restful}
Andrea Arcuri.
\newblock Restful api automated test case generation with evomaster.
\newblock {\em ACM Transactions on Software Engineering and Methodology (TOSEM)}, 28(1):1--37, 2019.

\bibitem{gross2012search}
Florian Gross, Gordon Fraser, and Andreas Zeller.
\newblock Search-based system testing: high coverage, no false alarms.
\newblock In {\em Proceedings of the 2012 International Symposium on Software Testing and Analysis}, pages 67--77, 2012.

\bibitem{mao2016sapienz}
Ke~Mao, Mark Harman, and Yue Jia.
\newblock Sapienz: Multi-objective automated testing for android applications.
\newblock In {\em Proceedings of the 25th international symposium on software testing and analysis}, pages 94--105, 2016.

\bibitem{sell2019empirical}
Leon Sell, Michael Auer, Christoph Fr{\"a}drich, Michael Gruber, Philemon Werli, and Gordon Fraser.
\newblock An empirical evaluation of search algorithms for app testing.
\newblock In {\em Testing Software and Systems: 31st IFIP WG 6.1 International Conference, ICTSS 2019, Paris, France, October 15--17, 2019, Proceedings 31}, pages 123--139. Springer, 2019.

\bibitem{cytron1991efficiently}
Ron Cytron, Jeanne Ferrante, Barry~K Rosen, Mark~N Wegman, and F~Kenneth Zadeck.
\newblock Efficiently computing static single assignment form and the control dependence graph.
\newblock {\em ACM Transactions on Programming Languages and Systems (TOPLAS)}, 13(4):451--490, 1991.

\bibitem{deiner2020search}
Adina Deiner, Christoph Fr{\"a}drich, Gordon Fraser, Sophia Geserer, and Niklas Zantner.
\newblock Search-based testing for scratch programs.
\newblock In {\em Search-Based Software Engineering: 12th International Symposium, SSBSE 2020, Bari, Italy, October 7--8, 2020, Proceedings 12}, pages 58--72. Springer, 2020.

\bibitem{gotz2022model}
Katharina G{\"o}tz, Patric Feldmeier, and Gordon Fraser.
\newblock Model-based testing of scratch programs.
\newblock In {\em 2022 IEEE Conference on Software Testing, Verification and Validation (ICST)}, pages 411--421. IEEE, 2022.

\bibitem{feldmeier2024combining}
Patric Feldmeier and Gordon Fraser.
\newblock Combining neuroevolution with the search for novelty to improve the generation of test inputs for games.
\newblock In {\em Proceedings of the 1st ACM International Workshop on Foundations of Applied Software Engineering for Games}, pages 14--19, 2024.

\bibitem{feldmeier2025many}
Patric Feldmeier, Katrin Schmelz, and Gordon Fraser.
\newblock Many-objective neuroevolution for testing games.
\newblock {\em arXiv preprint arXiv:2501.07954}, 2025.

\bibitem{arcuri2011practical}
Andrea Arcuri and Lionel Briand.
\newblock A practical guide for using statistical tests to assess randomized algorithms in software engineering.
\newblock In {\em Proceedings of the 33rd international conference on software engineering}, pages 1--10, 2011.

\bibitem{vargha2000critique}
Andr{\'a}s Vargha and Harold~D Delaney.
\newblock A critique and improvement of the cl common language effect size statistics of mcgraw and wong.
\newblock {\em Journal of Educational and Behavioral Statistics}, 25(2):101--132, 2000.

\bibitem{mann1947test}
Henry~B Mann and Donald~R Whitney.
\newblock On a test of whether one of two random variables is stochastically larger than the other.
\newblock {\em The annals of mathematical statistics}, pages 50--60, 1947.

\bibitem{offutt1996experimental}
A~Jefferson Offutt, Ammei Lee, Gregg Rothermel, Roland~H Untch, and Christian Zapf.
\newblock An experimental determination of sufficient mutant operators.
\newblock {\em ACM Transactions on Software Engineering and Methodology (TOSEM)}, 5(2):99--118, 1996.

\bibitem{kim2018guiding}
Jinhan Kim, Robert Feldt, and Shin Yoo.
\newblock Guiding deep learning system testing using surprise adequacy.
\newblock In {\em 2019 IEEE/ACM 41st International Conference on Software Engineering (ICSE)}, pages 1039--1049. IEEE, 2019.

\bibitem{rosenblatt1956remarks}
Richard~A Davis, Keh-Shin Lii, and Dimitris~N Politis.
\newblock Remarks on some nonparametric estimates of a density function.
\newblock In {\em Selected Works of Murray Rosenblatt}, pages 95--100. Springer, 2011.

\end{thebibliography}

%
%

\section*{Acknowledgments}
We would like to express our sincere gratitude to \href{https://scholar.google.com/citations?user=BXL5xZsAAAAJ&hl=en&oi=ao}{Patric Feldmeier} for his invaluable support during the execution and implementation setup of our approach. His previous work in the development of the NEATest algorithm served as a key foundation for our work. We greatly appreciate his contributions and collaboration throughout this research.

\section*{Author Contributions}

\textbf{Conceptualization:} Khizra Sohail; \textbf{Methodology:} Khizra Sohail, Atif Aftab Ahmed Jilani; \textbf{Validation:} Khizra Sohail; \textbf{Result analysis:} Khizra Sohail, Nigar Azhar Butt; \textbf{Visualization:} Khizra Sohail, Nigar Azhar Butt; \textbf{Supervision:} Atif Aftab Ahmed Jilani; \textbf{Writing – original draft:} Khizra Sohail, Nigar Azhar Butt; \textbf{Writing – review \& editing:} Khizra Sohail, Nigar Azhar Butt, Atif Aftab Jilani.

\section*{Data Availability}

 The data for framework is available at \url{https://github.com/khizra777/Neatest-Branch}.
 \\ The data for results is available at \url{https://github.com/khizra777/NEATEST-Branch-Evaluation}.

\section*{Funding} 
The authors report that there are no funding details to declare.

\section*{Competing Interests}
The authors declare that they have no competing interests related to this study.

\end{document}